\documentclass[aps,preprint,superscriptaddress,prb]{revtex4-2}
\usepackage{amsmath}
\usepackage{amssymb}
\usepackage{graphicx}
\usepackage[english]{babel}
\usepackage{bm}
\usepackage{mathrsfs}
\usepackage{hyperref}
\hypersetup{
    colorlinks=true,
    linkcolor=blue,
    citecolor=blue,      
    urlcolor=blue,
}

\newcommand{\op}[2][b]{\hat{#1}_{\mathbf{#2}}}
\newcommand{\opc}[2][b]{\hat{#1}^{\dagger}_{\mathbf{#2}}}

\newcommand{\omegaLO}{\omega_{\text{LO}}}
\newcommand{\figref}[1]{Fig. \ref{#1}}
\newcommand{\secref}[1]{Sec. \ref{#1}}
\newcommand{\tabref}[1]{Table \ref{#1}}
\newcommand{\tildeff}{\mathop{}\!\mathrm{d}}

\begin{document}

\title{Stability conditions for a large anharmonic bipolaron}

\author{Matthew Houtput}
\affiliation{Theory of Quantum Systems and Complex Systems, Universiteit Antwerpen, B-2000 Antwerpen, Belgium}

\author{Jacques Tempere}
\affiliation{Theory of Quantum Systems and Complex Systems, Universiteit Antwerpen, B-2000 Antwerpen, Belgium}

\begin{abstract}
A large polaron is a quasiparticle that consists of a nearly free electron interacting with the phonons of a material, whose lattice parameters are much smaller than the polaron scale. The electron-phonon interaction also leads to an attractive interaction between electrons, which can allow two polarons to pair up and form a bipolaron. It has been shown that large bipolarons can form in theory due to strong 1-electron-1-phonon coupling, but they have not been seen in real materials because the critical value of the required electron-phonon interaction is too large. Here, we investigate the effect of 1-electron-2-phonon coupling on the large bipolaron problem.

Starting from a generalization of the Fr\"ohlich Hamiltonian that includes both the standard 1-electron-1-phonon interaction as well as an anharmonic 1-electron-2-phonon interaction, we use the path integral method to find a semi-analytical upper bound for the bipolaron energy that is valid at all values of the Fr\"ohlich coupling strength $\alpha$. We find the bipolaron phase diagram and conditions for the bipolaron stability by comparing the bipolaron energy to the energy of two free polarons. The critical value of the Fr\"ohlich coupling strength $\alpha_{\text{crit}}$ is calculated as a function of the strength of the 1-electron-2-phonon interaction. The results suggest that large bipolaron formation is more likely in materials with significant 1-electron-2-phonon interaction as well as strong 1-electron-1-phonon interaction, such as strontium titanate.
\end{abstract}

\maketitle

\section{Introduction} \label{sec:Introduction}
An electron moving through an ionic lattice will interact with the phonons of that lattice, since both the electron and the ions are charged. This effect is known as electron-phonon interaction. If there is only one electron moving through the lattice, the electron will become dressed by the phonons, which leads to the polaron quasiparticle. The polaron was originally proposed in solids \cite{Landau1933, Landau1948}, but in general, an impurity interacting with a bosonic field is also known as a polaron in many contexts. Other examples include spin polarons \cite{Nagaev1974}, exciton polarons \cite{Verzelen2002}, ripplopolarons \cite{Tempere2003}, magnetic polarons \cite{Koepsell2019}, and the Bose polaron in ultracold gases \cite{Jorgensen2016,Shchadilova2016}. In solids, a distinction is made between ``large'' and ``small'' polarons, where the radius of the polaron wavefunction is much larger (or smaller, respectively) than the size of the unit cell of the crystal. 

While the single polaron problem is interesting on its own and has been extensively studied in the context of both many-body physics \cite{Frohlich1954, Feynman1955, Holstein1959, Devreese1972, Mishchenko2000, Alexandrov2010, Hahn2018, Klimin2020} and density functional theory \cite{Franchini2009, McKenna2012, Setvin2014, Sio2019, Franchini2021, LafuenteBartolome2022}, electron-phonon interaction becomes especially important when more than one electron is involved. Electron-phonon interaction causes an attractive interaction between electrons which allows them to form pairs: Bose-Einstein condensation of those pairs then leads to superconductivity. In a many-electron gas with weak electron-phonon coupling, the electrons can form Cooper pairs: this leads to conventional superconductivity, which is well described by Bardeen-Cooper-Schrieffer theory or Eliashberg theory \cite{Mahan2000, Bardeen1957, Eliashberg1960}.

In the limit of strong electron-phonon interaction, two electrons can form a bipolaron \cite{Emin1989, Verbist1991, Emin1992}. Contrarily to Cooper pair formation, a bipolaron can be understood with only two electrons instead of requiring the thermodynamic limit of many electrons \cite{Emin1989, Verbist1991}. In the past, bipolaron formation has been proposed as a possible pairing mechanism for unconventional high-$T_c$ superconductors like La$_2$CuO$_4$ or YBa$_2$Cu$_3$O$_7$ \cite{Schafroth1955, Alexandrov1981, Emin1992, Jourdan2003}. However, this proposal is controversial \cite{Chakraverty1998, Alexandrov1999}, largely because there is no experimental evidence of large bipolarons in 3D materials.

\begin{figure}
\centering
\includegraphics[width=8.6cm]{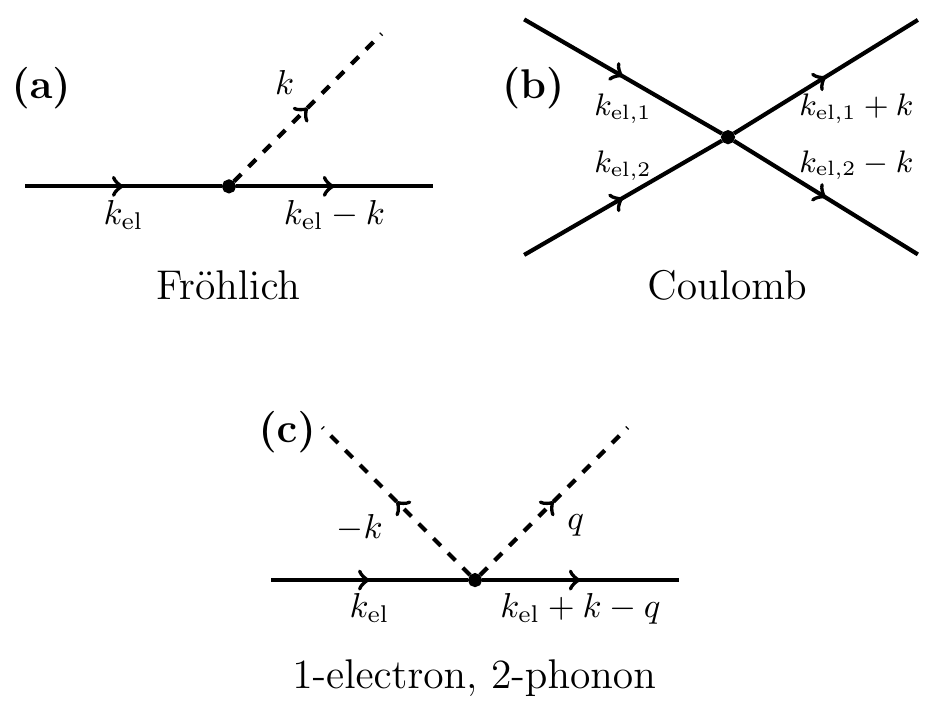}
\caption{\label{fig:HamRepresentation} Interactions between electrons (solid lines) and phonons (dashed lines) that are considered in this article. The previous treatment of bipolarons using path integrals \cite{Verbist1991} includes the Fr\"ohlich interaction in (\textbf{a}) and the Coulomb interaction in (\textbf{b}): this article also includes the 1-electron-2-phonon interaction shown in (\textbf{c}). }
\end{figure}

For large polarons in polar semiconductors, the electron-phonon coupling is usually well described by the Fr\"ohlich Hamiltonian \cite{Frohlich1954} and the strength of this coupling is fully determined by a single coupling parameter $\alpha$. Within the Fr\"ohlich model, a variational analysis using the path integral method \cite{Verbist1991} has shown that large bipolaron formation in 3D is possible above a critical value $\alpha > \alpha_{\text{crit}} = 6.8$. The Fr\"ohlich model is excellent for harmonic materials, and assumes that the electron-phonon coupling is linear as in \figref{fig:HamRepresentation}a. However, it fails when describing highly anharmonic superconductors, such as metallic hydrogen \cite{Ashcroft1968, Dias2017, Loubeyre2020}, high-pressure hydrides \cite{Drozdov2015, Somayazulu2019, Errea2015}, and quantum paraelectrics such as potassium tantalate or strontium titanate \cite{Collignon2019, Gastiasoro2020, Kumar2021}. The pairing mechanism in strontium titanate is the subject of ongoing research \cite{Gastiasoro2020}. One of the proposed pairings mechanisms is due to a 1-electron-2-phonon coupling to the soft transverse optical (TO) mode of strontium titanate \cite{Vandermarel2019, Feigelman2021}. The 1-electron-2-phonon interaction, depicted in \figref{fig:HamRepresentation}c, has been used to explain several other properties of strontium titanate, such as the $T^2$-behavior of the resistivity at low temperatures \cite{Kumar2021}.

In a recent paper \cite{Houtput2021}, it was shown that the 1-electron-2-phonon interaction of \figref{fig:HamRepresentation}c naturally appears alongside 3-phonon interaction when deriving the Fr\"ohlich Hamiltonian for an anharmonic material. It is possible to write down an analytical expression for the interaction strengths of the 1-electron-2-phonon interaction (\figref{fig:HamRepresentation}c) with longitudinal optical (LO) phonons \cite{Houtput2021}, suitable under the same conditions that are used for the Fr\"ohlich Hamiltonian \cite{Frohlich1954}. The goal of this paper is to investigate the stability of large bipolarons when this anharmonic 1-electron-2-phonon interaction is taken into account. Since only two phonons participate in this anharmonic interaction, the resulting Hamiltonian is still quadratic, and it is still possible to use the all-coupling path integral method used in \cite{Verbist1991}.

This paper is structured as follows. In \secref{sec:System}, the Hamiltonian and Lagrangian of the anharmonic bipolaron problem are introduced, and the phonons are eliminated in order to obtain an effective action functional. In \secref{sec:Model}, the variational bipolaron model system is introduced and discussed, and this model system is used to obtain a variational upper bound for the anharmonic bipolaron energy. This energy is used in \secref{sec:Stability} to investigate the region where bipolaron formation is possible. We conclude in \secref{sec:Conclusions}.

\section{The anharmonic bipolaron system} \label{sec:System}
\subsection{Extended Fr\"ohlich Hamiltonian}
In order to study the effect of 1-electron-2-phonon interaction on the bipolaron properties, we use the following extension to the Fr\"ohlich Hamiltonian that was derived in \cite{Houtput2021}, written in first quantization for the electrons and second quantization for the phonons:
\begin{align}
\hat{H} & = \hat{H}_{\text{el}} + \hat{H}_{\text{ph}} + \hat{H}_{\text{el-ph}}, \\
\hat{H}_{\text{el}} & = \sum_{i=1}^N \frac{\hat{\mathbf{p}}^2_{i}}{2 m_b} +  \frac{1}{2} \sum_{\mathbf{k} \neq \mathbf{0}} V^{(C)}_{\mathbf{k}} (\hat{\rho}_{\mathbf{k}} \hat{\rho}_{-\mathbf{k}} - N),  \label{HamEl} \\
\hat{H}_{\text{ph}} & = \sum_{\mathbf{k}} \hbar \omega_{\mathbf{k}} \left( \opc{k} \op{k} + \frac{1}{2} \right), \label{HamPh} \\
\hat{H}_{\text{el-ph}} & = \sum_{\mathbf{k} \neq \mathbf{0}} V^{(F)}_{\mathbf{k}} \left( \opc{k} + \op{-k} \right) \hat{\rho}_{-\mathbf{k}} + \frac{1}{2} \sum_{\mathbf{k} \neq \mathbf{k}' \neq \mathbf{0}} V^{(1)}_{\mathbf{k},\mathbf{k}'} \left( \opc{-k} + \op{k} \right)\left( \opc{k'} + \op{-k'} \right) \hat{\rho}_{\mathbf{k}-\mathbf{k}'}. \label{HamElPh}
\end{align}
Here $\opc{k}$ and $\op{k}$ represent the creation and annihilation operators of the LO phonons, $\hat{\mathbf{r}}_{i}$ and $\hat{\mathbf{p}}_{i}$ are the individual electron position and momentum operators, $\hat{\rho}_{\mathbf{k}} = \sum_{j=1}^N e^{i \mathbf{k} \cdot \hat{\mathbf{r}}_{j}}$ is the density operator of the electrons, and $N$ is the number of electrons (for a bipolaron, $N=2$).
This Hamiltonian is valid under the same conditions as the Fr\"ohlich Hamiltonian:  it applies to materials with cubic symmetry and two atoms in the unit cell, and it is assumed that the electrons only interact with the longitudinal optical phonon branch \cite{Houtput2021}. If there is an inversion center, it must hold that $V^{(1)}_{\mathbf{k},\mathbf{k}'} = 0$: therefore, it is assumed that the material does not have an inversion center. III-V semiconductors with the zincblende structure satisfy all of these constraints. Also note that due to the exclusion of the cases $\mathbf{k} = \mathbf{0}, \mathbf{k}' = \mathbf{0}$ and $\mathbf{k} = \mathbf{k}'$ in the sums of \eqref{HamEl}-\eqref{HamElPh}, one may assume that $V^{(F)}_{\mathbf{0}} = V^{(1)}_{\mathbf{0},\mathbf{k}} = V^{(1)}_{\mathbf{k},\mathbf{0}} = V^{(1)}_{\mathbf{k},\mathbf{k}} = 0$.

\begin{table}
\centering
\begin{tabular}{r|c|c|c|c}
 & $U$ & $\alpha$ & $\mathcal{T}_1$ & $\tilde{V}_0$ \\ \hline
\textbf{BN} & 4.070 & 0.973 & -0.00134 & 0.00121 \\
\textbf{BP} & 2.625 & 0.018 & -0.00085 & 0.00123 \\
\textbf{AlN} & 4.566 & 1.492 & -0.00069 & 0.00100 \\
\textbf{AlP} & 3.638 & 0.561 & 0.00050 & 0.00092 \\
\end{tabular}
\caption{\label{tab:MaterialParameters} Values for the model parameters $U$, $\alpha$, $\mathcal{T}_1$ and $\tilde{V}_0$ in \eqref{VCoulomb}-\eqref{VAnhar1} for the lightest III-V semiconductors in the zincblende structure (space group $F\bar{4}3m$). The values in this table are taken or calculated from Table I in \cite{Houtput2022}.}
\end{table}
The Hamiltonian is fully described by the phonon dispersion $\omega_{\mathbf{k}}$ and the momentum-dependent interaction strengths $V^{(F)}_{\mathbf{k}}$, $V^{(C)}_{\mathbf{k}}$ and $V^{(1)}_{\mathbf{k},\mathbf{k}'}$ of the processes in \figref{fig:HamRepresentation}. We assume a single dispersionless LO phonon branch as in the Fr\"ohlich model \cite{Frohlich1954}, and we use the interaction strengths from \cite{Houtput2021}:
\begin{align}
\omega_{\mathbf{k}} & = \omegaLO, \label{PhononWK} \\
V^{(C)}_{\mathbf{k}} & = \hbar \omegaLO \frac{4 \pi \sqrt{2} U}{V} \sqrt{\frac{\hbar}{2 m_b \omegaLO}} \frac{1}{|\mathbf{k}|^2},
\label{VCoulomb} \\
V^{(F)}_{\mathbf{k}} & = \hbar \omegaLO \sqrt{\frac{4\pi\alpha}{V} } \left(\frac{\hbar}{2m_b \omegaLO} \right)^{\frac{1}{4}} \frac{1}{|\mathbf{k}|}, \label{VFrohlich} \\
V^{(1)}_{\mathbf{k},\mathbf{k}'} & = -i \hbar \omegaLO \frac{\sqrt{4\pi \alpha} \mathcal{T}_1}{V} \frac{\hbar}{2m_b \omegaLO} |\varepsilon_{ijl}| \frac{k_i (k_j-k'_j) k'_l}{|\mathbf{k}||\mathbf{k}-\mathbf{k}'|^2|\mathbf{k}|}. \label{VAnhar1}
\end{align}
Here $|\varepsilon_{ijl}|$ is the absolute value of the Levi-Civita tensor. All results can be plotted in terms of four dimensionless material parameters, which are assumed to be known: the strength of the Coulomb interaction $U$, the strength of the Fr\"ohlich interaction $\alpha$ \cite{Frohlich1954}, the strength of the 1-electron-2-phonon interaction $\mathcal{T}_1$, and the dimensionless size of the unit cell $\tilde{V}_0 = V_0/\left( \frac{\hbar}{2 m_b \omegaLO} \right)^{\frac{3}{2}}$ which appears to regularize divergent momentum integrals \cite{Houtput2021, Houtput2022}. Every material has a single value of $U$, $\alpha$, $\mathcal{T}_1$, and $\tilde{V}_0$: these values can be found in \tabref{tab:MaterialParameters}. $U$ and $\alpha$ can be expressed in terms of the low- and high-frequency transverse dielectric constants $\varepsilon_0$ and $\varepsilon_{\infty}$:
\begin{align}
U & = \frac{1}{\sqrt{2} \hbar \omegaLO} \frac{e^2}{4 \pi \varepsilon_{\text{vac}} \varepsilon_{\infty}} \sqrt{\frac{2 m_b \omegaLO}{\hbar}}, \label{UDef} \\
\alpha & = \frac{1}{\sqrt{2}} \left(1 - \frac{\varepsilon_{\infty}}{\varepsilon_{0}} \right) U. \label{alphaDef}
\end{align}
Given a value for $U$, $\alpha$, and $\mathcal{T}_1$, the interaction strengths \eqref{VCoulomb}-\eqref{VFrohlich} and by extension the anharmonic bipolaron Hamiltonian \eqref{HamEl}-\eqref{HamElPh} are uniquely determined. In this article, we use the path integral method to find the ground state energy of this Hamiltonian.

\subsection{Physical conditions on the material parameters} \label{sec:PhysicalConditions}
For a physical material, the parameters $U$, $\alpha$, $\mathcal{T}_1$ and $\tilde{V}_0$ cannot take on completely arbitrary values. For example, since $\varepsilon_0 > \varepsilon_{\infty} > 0$ for any physical material, it can be seen from equations \eqref{UDef}-\eqref{alphaDef} that the values of $U$ and $\alpha$ must satisfy the following physical condition \cite{Verbist1991}:
\begin{equation} \label{PhysicalCondition}
U > \sqrt{2} \alpha \geq 0.
\end{equation}
For the Fr\"ohlich bipolaron ($\mathcal{T}_1 = 0$), this condition is sufficient \cite{Verbist1991}. However, when 1-electron-2-phonon interaction is included, one needs to be more careful. In this section, a new physical condition is derived by investigating the transverse and longitudinal dielectric functions of the Hamiltonian \eqref{HamEl}-\eqref{HamElPh}.

The transverse dielectric function represents the response of the system to an oscillating electric field $\mathbf{E}(\omega)$ \cite{Kittel2004}. It has been calculated in \cite{Houtput2021} for the anharmonic Hamiltonian \eqref{HamEl}-\eqref{HamPh}, and can be written in terms of $U$ and $\alpha$ as:
\begin{equation} \label{EpsilonTransverse}
\varepsilon_T(\omega) = \varepsilon_{\infty} \frac{\omega^2 - \omegaLO^2}{\omega^2 - \omegaLO^2 \left(1- \frac{\sqrt{2} \alpha}{U}\right)}.
\end{equation}
The transverse dielectric function of the anharmonic Hamiltonian is independent of $\mathcal{T}_1$ and is therefore the same as the polariton-type dielectric function of the Fr\"ohlich Hamiltonian \cite{Ashcroft1976, Kittel2004}. Requiring that $\varepsilon_0 := \varepsilon_T(0) > 0$ immediately leads to condition \eqref{PhysicalCondition}.

\begin{figure}
\centering
\includegraphics[width=8.6cm]{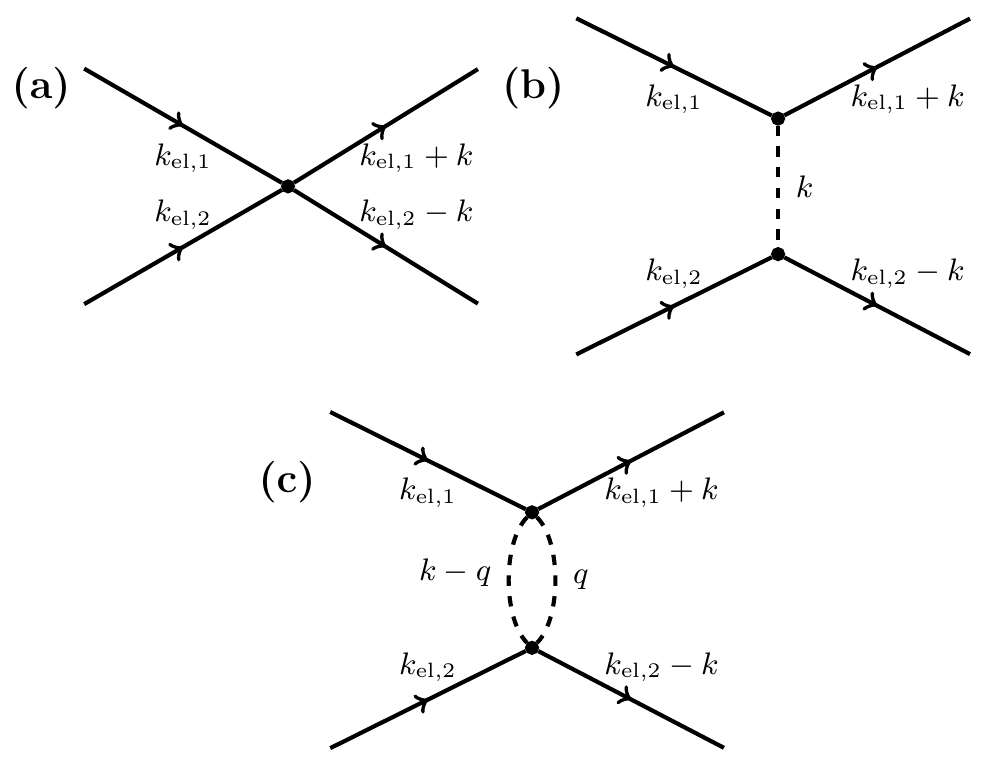}
\caption{\label{fig:ScreenedCoulomb} The three different interaction processes between two electrons that contribute to the longitudinal dielectric function $\varepsilon_L(\omega)$: (\textbf{a}) Direct Coulomb interaction \cite{Mahan2000}, (\textbf{b}) Exchanging one phonon via the Fr\"ohlich interaction \cite{Frohlich1954, Mahan2000}, and (\textbf{c}) Exchanging two phonons via the 1-electron-2-phonon interaction \cite{Houtput2021}. $k_{\text{el}}$, $k$ and $q$ represent four-momenta of the form $k = (\omega,\mathbf{k})$.}
\end{figure}
The longitudinal dielectric function represents the screening of the Coulomb interaction between two electrons due to the exchange of phonons \cite{Mahan2000}. \figref{fig:ScreenedCoulomb} shows the three different ways in which two electrons can interact, which now includes a process involving 1-electron-2-phonon interactions. The longitudinal dielectric function $\varepsilon_L(\omega)$ is defined such that the sum of all these processes can be written as a single interaction vertex \cite{Mahan2000}:
\begin{equation} \label{ScreenedInteractionVertex}
V^{(C,\text{screened})}_{\mathbf{k}}(\omega) := \frac{e^2}{V \varepsilon_{\text{vac}} \varepsilon_L(\omega)} \frac{1}{|\mathbf{k}|^2}.
\end{equation}
The diagrams in \figref{fig:ScreenedCoulomb} have been drawn including the incoming and outgoing electron lines. However, for the calculation of the interaction vertex \eqref{ScreenedInteractionVertex}, these four lines should be ignored. Interpreting the remaining parts of the diagrams in \figref{fig:ScreenedCoulomb} using the Feynman rules in \cite{Goldberg1985, Houtput2021} and adding up the three diagrams yields the following expression for the longitudinal dielectric function:
\begin{align}
\frac{1}{\varepsilon_L(\omega)} & = \frac{1}{\varepsilon_{\infty}} + \frac{V \varepsilon_{\text{vac}} \varepsilon_{\infty} |\mathbf{k}|^2}{e^2 \hbar} \left|V^{(F)}_{\mathbf{k}}\right|^2 D_0(\omega) \nonumber \\
& + \frac{i V \varepsilon_{\text{vac}} \varepsilon_{\infty} |\mathbf{k}|^2}{2 e^2 \hbar} \sum_{\mathbf{q}} \left|V^{(1)}_{-\mathbf{k}+\mathbf{q},\mathbf{q}}\right|^2 \int_{-\infty}^{+\infty} D_0(\omega-\nu) D_0(\nu) \frac{\tildeff \nu}{2\pi}, \label{epsInt1}
\end{align}
where $D_0(\omega)$ is the free phonon Green's function \cite{Mahan2000, Houtput2021}. The sums and integrals in expression \eqref{epsInt1} were calculated in \cite{Houtput2021}. A straightforward calculation yields:
\begin{equation} \label{EpsilonLongitudinal}
\frac{1}{\varepsilon_L(\omega)} = \frac{1}{\varepsilon_{\infty}} \left[1 + \frac{\sqrt{2} \alpha}{U} \left(\frac{\omegaLO^2}{\omega^2 - \omegaLO^2} + \frac{ \mathcal{T}_1^2}{15 \tilde{V}_0} \frac{4 \omegaLO^2}{\omega^2 - 4 \omegaLO^2} \right) \right].
\end{equation}
In general, the longitudinal dielectric function is therefore not equal to the transverse dielectric function \eqref{EpsilonTransverse}. Taking the high-frequency limit of \eqref{EpsilonLongitudinal} shows that $\varepsilon_L(+\infty)$ is equal to $\varepsilon_{\infty}$. However, the low-frequency limit gives that $\varepsilon_L(0) \neq \varepsilon_0$:
\begin{equation} \label{EpsilonLongitudinalConstant}
\varepsilon_{L}(0) = \frac{\varepsilon_{\infty}}{1 - \frac{\sqrt{2} \alpha}{U} \left(1 + \frac{\mathcal{T}_1^2}{15 \tilde{V}_0} \right)}.
\end{equation}
Therefore, we must explicitly specify that $\varepsilon_0$ represents the \emph{transverse} dielectric function at zero frequency. More importantly, for any physical material it must hold that $\varepsilon_L(0) > 0$: otherwise, the screened Coulomb interaction becomes attractive, and the electron gas is unstable.  Therefore, we obtain the following physical condition on $U$, $\alpha$, $\mathcal{T}_1$, and $\tilde{V}_0$:
\begin{equation} \label{PhysicalConditionAnharmonic}
U > \sqrt{2} \alpha \left(1 + \frac{\mathcal{T}_1^2}{15 \tilde{V}_0} \right) > 0.
\end{equation}
If \eqref{PhysicalConditionAnharmonic} is satisfied, then so is the other physical condition \eqref{PhysicalCondition}: therefore, it is sufficient to only take condition \eqref{PhysicalConditionAnharmonic} into account for the anharmonic bipolaron problem.

Note that the distinction between longitudinal and transverse dielectric functions is unnecessary when discussing the Fr\"ohlich Hamiltonian. Indeed, when $\mathcal{T}_1 = 0$, the longitudinal dielectric function \eqref{EpsilonLongitudinal} reduces to the transverse dielectric function \eqref{EpsilonTransverse}, and the two physical conditions \eqref{PhysicalCondition} and \eqref{PhysicalConditionAnharmonic} are the same. This explains why it is usually not specified in the literature whether $\varepsilon_0$ represents the longitudinal or transverse dielectric constant \cite{Frohlich1954}, and why this distinction is not made in \cite{Verbist1991} to determine the physical condition \eqref{PhysicalCondition}.

\subsection{Imaginary time Lagrangian and effective action functional}
The path integral method was first applied to the polaron problem by Feynman \cite{Feynman1955} and has since been used in many other treatments of polarons \cite{Feynman1962, Gerlach1987, Rosenfelder2001, Ichmoukhamedov2019}. The basic idea behind the method is to write the partition sum $Z$ at inverse temperature $\beta$ as a double path integral over the electron coordinates $\mathbf{r}_i(\tau)$ and the phonon coordinates $q_{\mathbf{k}}(\tau)$:
\begin{equation} \label{Zdef}
Z = \int \mathcal{D}\mathbf{r}_{i}(\tau) \int \mathcal{D}q_{\mathbf{k}}(\tau) \exp\left(-\frac{1}{\hbar} \int_0^{\hbar \beta} L\left(q_{\mathbf{k}}(\tau),\dot{q}_{\mathbf{k}}(\tau), \mathbf{r}_{i}(\tau),\dot{\mathbf{r}}_{i}(\tau) \right) \tildeff\tau \right),
\end{equation}
where $L\left(q_{\mathbf{k}}(\tau),\dot{q}_{\mathbf{k}}(\tau), \mathbf{r}_{i}(\tau),\dot{\mathbf{r}}_{i}(\tau) \right)$ is the classical Lagrangian of the system in imaginary time. For the anharmonic polaron Hamiltonian \eqref{HamEl}-\eqref{HamElPh}, the imaginary time Lagrangian has already been calculated \cite{Houtput2021}:
\begin{align}
& L\left(q_{\mathbf{k}}(\tau),\dot{q}_{\mathbf{k}}(\tau), \mathbf{r}_{i}(\tau),\dot{\mathbf{r}}_{i}(\tau) \right) \nonumber \\
 & = \sum_{i=1}^N \frac{m_b}{2} [\dot{\mathbf{r}_{i}}(\tau)]^2 + \frac{1}{2} \sum_{\mathbf{k}} V^{(C)}_{\mathbf{k}} (\rho_{\mathbf{k}}(\tau)\rho_{-\mathbf{k}}(\tau)-N) + \underset{\mathbf{k}}{\sum} \frac{m_{\text{ph}}}{2} \left( |\dot{q}_{\mathbf{k}}(\tau)|^2 + \omega_{\mathbf{k}}^2 |q_{\mathbf{k}}(\tau)|^2 \right) \nonumber \\
 & + \text{Re}\left[
 \sum_{\mathbf{k}} \sqrt{\frac{2 m_{\text{ph}} \omega_{\mathbf{k}}}{\hbar}} V^{(F)}_{\mathbf{k}} \rho_{\mathbf{k}}(\tau) q_{\mathbf{k}}(\tau) \right] +\text{Re}\left[ \underset{\mathbf{k},\mathbf{k}'}{\sum} \frac{2 m_{\text{ph}} \sqrt{\omega_{\mathbf{k}} \omega_{\mathbf{k}'}}}{\hbar} V^{(1)}_{\mathbf{k},\mathbf{k}'} \rho_{\mathbf{k}-\mathbf{k}'}(\tau) q_{\mathbf{k}}(\tau) q^*_{\mathbf{k}'}(\tau)
  \right], \label{Lagrangian}
\end{align}
where $m_{\text{ph}}$ is an arbitrary phonon mass that does not influence the results, and:
\begin{equation} \label{rhoDef}
\rho_{\mathbf{k}}(\tau) = \sum_{j=1}^N e^{i \mathbf{k} \cdot \mathbf{r}_j(\tau)}
\end{equation}
is the classical analog of the electron density operator. The only difference with the Lagrangian in \cite{Houtput2021} is the Coulomb interaction term, which carries over directly from the Hamiltonian.

Since the Lagrangian \eqref{Lagrangian} is quadratic in the phonon coordinates $q_{\mathbf{k}}(\tau)$, the phonon path integral in \eqref{Zdef} is a Gaussian integral which may be performed exactly. This integral was performed in \cite{Houtput2021} for a single electron. However, since the calculation does not assume a specific expression for $\rho_{\mathbf{k}}(\tau)$ \cite{HoutputPhD}, and the additional Coulomb interaction does not depend on $q_{\mathbf{k}}(\tau)$, the result in \cite{Houtput2021} is valid for many electrons as long as we add the Coulomb interaction afterwards. Therefore, the partition sum $Z$ can be written as a path integral over only the electron coordinates:
\begin{equation}\label{ZPart}
Z = \left(\prod_{\mathbf{k}} \frac{1}{2\sinh\left(\frac{\hbar \beta \omega_{\mathbf{k}}}{2}\right)}\right) \int \mathcal{D}\mathbf{r}_{i}(\tau) \exp\left(-\frac{1}{\hbar} S_{\text{eff}}[\mathbf{r}_{i}(\tau)] \right). 
\end{equation}
The effective action functional for the electrons $S_{\text{eff}}[\mathbf{r}_{i}(\tau)]$ can be written as the free electron kinetic energy $S_{\text{free}} := \sum_{i=1}^N \int_0^{\hbar \beta} \frac{1}{2} m_b [\dot{\mathbf{r}}_{i}(\tau)]^2 \tildeff\tau$, plus several interaction terms:
\begin{equation} \label{SEffGeneral}
S_{\text{eff}}[\mathbf{r}_{i}(\tau)] = S_{\text{free}}[\mathbf{r}_i(\tau)] - \hbar O_C[\mathbf{r}_{i}(\tau)] -\hbar \sum_{n=0}^{+\infty} (-1)^n O_n[\mathbf{r}_{i}(\tau)] -\hbar \sum_{n=1}^{+\infty} \frac{(-1)^n}{n} \tilde{O}_n[\mathbf{r}_{i}(\tau)].
\end{equation}
The effective action functional \eqref{SEffGeneral} consists of a Coulomb interaction term $O_C$ and two infinite series containing the terms $O_n$ and $\tilde{O}_n$: each of these terms is of order $\mathcal{T}_1^n$ in the anharmonic interaction. The general expressions for $O_n$ and $\tilde{O}_n$ can be found in \cite{Houtput2021}. In \cite{Houtput2021}, the energy of a single polaron was calculated under the assumption that the anharmonic interaction is weak ($\mathcal{T}_1 \ll 1$), and only the first nonzero terms in the infinite series $O_0$ and $\tilde{O}_2$ were kept. In \secref{sec:Model} of this article, we must make the same assumptions in order to proceed analytically. Under this approximation, the effective action becomes:
\begin{equation} \label{SEff}
S_{\text{eff}}[\mathbf{r}_{i}(\tau)] \approx \int_0^{\hbar \beta} \sum_{j=1}^N \frac{1}{2} m_b [\dot{\mathbf{r}}_{j}(\tau)]^2 \tildeff\tau - \hbar O_C[\mathbf{r}_{i}(\tau)] -\hbar O_0[\mathbf{r}_{i}(\tau)] - \frac{\hbar}{2} \tilde{O}_2[\mathbf{r}_{i}(\tau)],
\end{equation}
where $O_C$ carries over directly from the Coulomb repulsion term in \eqref{Lagrangian}, and $O_0$ and $\tilde{O}_2$ are given by the following expressions \cite{Houtput2021}:
\begin{align}
O_C[\mathbf{r}_{i}(\tau)] := & \frac{1}{2 \hbar} \sum_{\mathbf{k}} \int_0^{\hbar \beta} V^{(C)}_{\mathbf{k}} (\rho^*_{\mathbf{k}}(\tau) \rho_{\mathbf{k}}(\tau) - N) \tildeff\tau, \label{OCExpr} \\
O_0[\mathbf{r}_i(\tau)] := & \frac{1}{2 \hbar^2} \sum_{\mathbf{k}} \int_0^{\hbar \beta} \int_0^{\hbar \beta} |V^{(F)}_{\mathbf{k}}|^2 \rho^*_{\mathbf{k}}(\tau)\rho_{\mathbf{k}}(\sigma) G_{\omegaLO}(\tau-\sigma)  \tildeff\tau  \tildeff\sigma, \label{O0Expr} \\
\tilde{O}_2[\mathbf{r}_i(\tau)] := & \frac{1}{2\hbar^2} \sum_{\mathbf{k}_1, \mathbf{k}_2} \int_0^{\hbar \beta}  \int_0^{\hbar \beta} \left|V^{(1)}_{\mathbf{k}_1,\mathbf{k}_2}\right|^2 \rho^*_{\mathbf{k}_2-\mathbf{k}_1}(\tau) \rho_{\mathbf{k}_2-\mathbf{k}_1}(\sigma) G_{\omegaLO}(\tau-\sigma)^2 \tildeff\tau \tildeff\sigma. \label{O2TildeExpr}
\end{align}
In these expressions, the dimensionless phonon Green's function $G_{\omega}(\tau)$ is defined as:
\begin{align} \label{PhononGreensFunction}
G_{\omega}(\tau) & := \frac{\cosh\left[\omega\left(\frac{\hbar \beta}{2}-|\tau|\right)\right]}{\sinh\left(\frac{\hbar \beta \omega}{2}\right)}. & &(-\hbar \beta < \tau < \hbar \beta).
\end{align}
These interaction terms are no longer local in time: the electrons can interact with the past image of themselves or the other electron. This memory effect is introduced because the phonons were eliminated from the description of the bipolaron. Within the weakly anharmonic approximation, the effective bipolaron action functional $S_{\text{eff}}$ is a sum of the kinetic energy and three independent interaction terms. The first interaction term $O_C$ represents the direct Coulomb repulsion between the two electrons. The second interaction term $O_0$ represents a retarded interaction due to the Fr\"ohlich interaction in the absence of anharmonicity. The third interaction term $\tilde{O}_2$ represents the dominant anharmonic retarded interaction: if $\tilde{O}_2$ is ignored, the effective action functional reduces to the harmonic expression in \cite{Verbist1991}.

We note that for the calculation of the path integral over the electrons in \eqref{ZPart}, it must be specified whether the electrons are distinguishable or not. To calculate the path integral over indistinguishable electrons, it is necessary to sum over all possible permutations of the end points of the those electrons \cite{Lemmens1996, Ichmoukhamedov2021_1, Ichmoukhamedov2021_2}, including a sign for odd permutations. For the remainder of the article, it is assumed that the electrons are distinguishable, for example because they have opposite spin: this corresponds to a bipolaron in the singlet state. This assumption is also made in \cite{Verbist1991}.

\section{The quadratic model system} \label{sec:Model}
\subsection{The Jensen-Feynman inequality}
In principle, the ground state energy of the bipolaron can be obtained by calculating the path integral for the partition sum $Z$ in \eqref{ZPart} with the effective action \eqref{SEffGeneral}, calculating the free energy $F = -\frac{1}{\beta} \text{ln}(Z)$, and taking the $\beta \rightarrow +\infty$ limit. However, the path integral \eqref{ZPart} is too complicated to evaluate analytically, even for one polaron in the harmonic approximation \cite{Feynman1955}. Therefore, we use the Jensen-Feynman inequality \cite{Feynman1955, Kleinert2009} to calculate an upper bound for the free energy. For any model action functional $S_0[\mathbf{r}_i(\tau)]$, the following inequality holds:
\begin{equation} \label{JensenFeynmanGeneral}
F \leq F_0 + \frac{1}{\hbar \beta} \left\langle S_{\text{eff}} - S_0 \right \rangle_0,
\end{equation}
where $F_0$ is the free energy of the model system and $\langle\rangle_0$ represents an expectation value with respect to the model action $S_0$, as defined in appendix \ref{sec:AppModel}. Both of these can always be calculated if the model action is quadratic in the electron coordinates. Equation \eqref{JensenFeynmanGeneral} can be used as a variational principle: The model action $S_0$ is chosen with several variational parameters in such a way that it mimics the original action.

The general effective action \eqref{SEffGeneral} consists of a kinetic energy term and several interaction terms. Therefore, it is possible to isolate the kinetic contribution $F_{\text{kin}}$ and interaction contribution $F_{\text{int}}$ of the free energy upper bound:
\begin{align}
F & \leq F_{\text{kin}} - F_{\text{int}}, \label{FBound} \\
F_{\text{kin}} & := F_0 - \frac{1}{\hbar \beta} \langle S_0 - S_{\text{free}} \rangle_0, \label{Fkin} \\
F_{\text{int}} & := \frac{1}{\beta} \langle O_C \rangle_0 + \frac{1}{\beta} \sum_{n=0}^{+\infty} (-1)^n \langle O_n \rangle_0 + \frac{1}{\beta} \sum_{n=1}^{+\infty} \frac{(-1)^n}{n} \langle \tilde{O}_n \rangle_0. \label{Fint}
\end{align}
The kinetic part $F_{\text{kin}}$ only depends on the choice of the model action and will be calculated exactly in \secref{sec:ModelAction}. At weak anharmonicity $\mathcal{T}_1 \ll 1$, the interaction part can be approximated by using expression \eqref{SEff} for the effective action functional:
\begin{equation} \label{FintApprox}
F_{\text{int}} \approx \frac{1}{\beta} \langle O_C \rangle_0 + \frac{1}{\beta}  \langle O_0 \rangle_0 + \frac{1}{2\beta} \langle \tilde{O}_2 \rangle_0.
\end{equation}
For this approximation, all of the higher order terms were discarded: these are $\langle O_n \rangle$ with $n\geq 1$ and $\langle \tilde{O}_n \rangle$ with $n \geq 3$. It can be shown \cite{HoutputPhD} that for the choices \eqref{VFrohlich}-\eqref{VAnhar1} for the interaction strengths, the odd order terms are all zero ($\langle O_{2n+1} \rangle = \langle \tilde{O}_{2n+1} \rangle = 0$) and the even order terms are all positive ($\langle O_{2n} \rangle \geq 0$ and $\langle \tilde{O}_{2n} \rangle  \geq 0$). Therefore, even with the approximate expression \eqref{FintApprox}, the variational upper bound in \eqref{FBound} still holds.

\subsection{Choice of the model action} \label{sec:ModelAction}
The model action $S_0[\mathbf{r}_1(\tau),\mathbf{r}_2(\tau)]$ should be chosen in such a way that it mimics the effective action functional \eqref{SEff} of the bipolaron, but it should also be quadratic in the electron coordinates so that its free energy $F_0$ and expectation values $\langle \rangle_0$ can be calculated exactly. We will choose a general quadratic model action, based on the model actions in \cite{Verbist1991, Rosenfelder2001, Ichmoukhamedov2021_1}:
\begin{align} \label{ModelAction}
& S_0[\mathbf{r}_{1}(\tau),\mathbf{r}_{2}(\tau)] = S_{\text{free}}[\mathbf{r}_{1}(\tau),\mathbf{r}_{2}(\tau)] \\
& \hspace{10pt} + \frac{m_b}{2} \int_0^{\hbar \beta} \int_0^{\hbar \beta} \left(
 \frac{1}{2} f(\tau-\sigma) \overset{2}{\underset{j=1}{\sum}} \left[\mathbf{r}_{j}(\tau)-\mathbf{r}_{j}(\sigma) \right]^2 + g(\tau-\sigma) \left[\mathbf{r}_{1}(\tau) - \mathbf{r}_{2}(\sigma) - \mathbf{a} \right]^2 \right) \tildeff\tau \tildeff\sigma.
\end{align}
This model action represents two electrons at an equilibrium distance $\mathbf{a}$ of each other. Just like the interactions \eqref{O0Expr}-\eqref{O2TildeExpr} in the effective action functional \eqref{SEff}, the electrons interact with themselves and each other in the past: in \eqref{ModelAction}, these interactions are determined by the memory functions $f(\tau)$ and $g(\tau)$, respectively. $\mathbf{a}$, $f(\tau)$, and $g(\tau)$ are the variational parameters of the model. It is assumed that $f(\tau)$ and $g(\tau)$ are defined on the interval $[0,\hbar \beta]$, and are periodically extended outside this interval. Without loss of generality, $f(\tau)$ and $g(\tau)$ can be chosen to be symmetric around $\tau = \hbar \beta/2$. The Coulomb repulsion between the electrons is also approximated as a quadratic repulsion around the equilibrium displacement $\mathbf{a}$, which is a significant drawback of this model. Note that the model action used in \cite{Verbist1991}, where the electrons are coupled to each other and fictitious phonon masses by springs, is a special case of \eqref{ModelAction} for a specific choice of $f(\tau)$ and $g(\tau)$. This means that the variational upper bound obtained with \eqref{ModelAction} will always be closer to the true free energy than the one obtained with the model system in \cite{Verbist1991}, though just like for the Fr\"ohlich polaron \cite{Rosenfelder2001}, the difference turns out to be negligible \cite{HoutputPhD}.

It is usually more convenient to write the electron paths $\mathbf{r}_j(\tau)$ in terms of their Fourier-Matsubara coefficients $\mathbf{r}_j(\omega_n)$, where $\omega_n$ are the bosonic Matsubara frequencies:
\begin{align}
\mathbf{r}_j(\tau) & = \sum_{n \in \mathbb{Z}} \mathbf{r}_j(\omega_n) e^{i \omega_n \tau} \label{rjFourierMatsubara}, \\
\mathbf{r}_j(\omega_n) & := \frac{1}{\hbar \beta} \int_0^{\hbar \beta} \mathbf{r}_j(\tau) e^{-i \omega_n \tau} d\tau, \label{rjRealSpace} \\
\omega_n & := \frac{2\pi n}{\hbar \beta}. \label{MatsubaraFrequencies}
\end{align}
Then, the model action can be written in terms of two profile functions $A_+(\omega_n)$ and $A_-(\omega_n)$, \cite{Rosenfelder2001}:
\begin{equation} \label{ModelActionFourier}
S_0 = \frac{m_b}{4} \hbar \beta \sum_{n \in \mathbb{Z}} \left[ \omega_n^2 A_+(\omega_n) \left|\mathbf{r}_{1}(\omega_n) + \mathbf{r}_{2}(\omega_n)\right|^2 + \omega_n^2 A_-(\omega_n) \left|\mathbf{r}_{1}(\omega_n) - \mathbf{r}_{2}(\omega_n) - \mathbf{a} \delta_{n,0}\right|^2 \right].
\end{equation}
The discrete set of values $A_{\pm}(\omega_n)$ can be used instead of $f(\tau)$ and $g(\tau)$ as variational parameters. Indeed, using \eqref{rjFourierMatsubara} in \eqref{ModelAction} and comparing with \eqref{ModelActionFourier} shows that the profile functions can be written in terms of $f(\tau)$ and $g(\tau)$ as follows:
\begin{align}
A_+(\omega) & = 1 + \frac{4}{\omega^2} \int_0^{\frac{\hbar \beta}{2}} \sin^2\left(\frac{\omega \tau}{2} \right) f(\tau) d\tau + \frac{4}{\omega^2} \int_0^{\frac{\hbar \beta}{2}} \sin^2\left(\frac{\omega \tau}{2} \right) g(\tau) d\tau, \label{ApDef} \\
A_-(\omega) & = 1 + \frac{4}{\omega^2} \int_0^{\frac{\hbar \beta}{2}} \sin^2\left(\frac{\omega \tau}{2} \right) f(\tau) d\tau + \frac{4}{\omega^2} \int_0^{\frac{\hbar \beta}{2}} \cos^2\left(\frac{\omega \tau}{2} \right) g(\tau) d\tau. \label{AmDef}
\end{align}
Both of these profile functions approach $1$ as $\omega \rightarrow +\infty$. In the low-frequency limit, $A_+(0)$ remains constant, but $A_-(\omega)$ diverges as $1/\omega^2$:
\begin{equation} \label{AmLimit}
\lim_{\omega \rightarrow 0} \omega^2 A_-(\omega) = 2 \int_0^{\hbar \beta} g(\tau) d\tau := 2 \hbar \beta g_0,
\end{equation}
where $g_0$ is the zeroth coefficient of the Fourier-Matsubara series of $g(\tau)$. It will appear in our intermediary results before the limit $\beta \rightarrow +\infty$ is taken. Since it can be calculated from $A_-(\omega)$, it should be seen as the variational parameter corresponding to $A_-(0)$.
\subsection{Expectation values with respect to the model action}
Because the model action functional \eqref{ModelAction} is quadratic, its free energy $F_0$ and expectation values $\langle \rangle_0$ can be calculated exactly in terms of the profile functions. In appendix \ref{sec:AppModel}, it is shown that:
\begin{align}
F_0 & = 2 F_{\text{free}} + \frac{3}{\beta} \sum_{n=1}^{+\infty} \left\{ \text{ln}[A_+(\omega_n)] + \text{ln}[A_-(\omega_n)] \right\} + \frac{1}{\beta} \text{ln}\left[V \left( \frac{m g_0}{\pi \hbar \beta^2} \right)^{\frac{3}{2}} \right], \label{FreeEnergyModel} \\
\frac{1}{\hbar \beta} \langle S_0 - S_{\text{free}} \rangle_0 & = \frac{3}{\beta} \left[\frac{1}{2} + \sum_{n=1}^{+\infty} \left(1 - \frac{1}{A_+(\omega_n)}\right) + \sum_{n=1}^{+\infty} \left(1 - \frac{1}{A_-(\omega_n)}\right) \right]. \label{ModelActionExpValue}
\end{align}
Here $F_{\text{free}} = - \frac{1}{\beta} \text{ln}\left[V \left(\frac{m_b}{2\pi \hbar^2 \beta} \right)^{\frac{3}{2}} \right]$ is the free energy of one free electron, and $V$ is the volume of the system. These two quantities are enough to calculate the kinetic part \eqref{Fkin} of the free energy upper bound:
\begin{align}
F_{\text{kin}} & = 2 F_{\text{free}} + \frac{1}{\beta} \text{ln}\left[V \left( \frac{m g_0}{\pi \hbar \beta^2} \right)^{\frac{3}{2}} \right] - \frac{3}{2\beta} \nonumber \\
& + \frac{3}{\beta} \sum_{n=1}^{+\infty} \left\{ \text{ln}[A_+(\omega_n)] + \frac{1}{A_+(\omega_n)} - 1 + \text{ln}[A_-(\omega_n)]  + \frac{1}{A_-(\omega_n)} - 1 \right\}. \label{FkinFinal}
\end{align}

In order to calculate the interaction free energy \eqref{FintApprox}, the expectation values of expressions \eqref{OCExpr}, \eqref{O0Expr}, and \eqref{O2TildeExpr} are required:
\begin{align}
\langle O_C \rangle_0 := & \frac{1}{2\hbar} \sum_{\mathbf{k}} \int_0^{\hbar \beta} V^{(C)}_{\mathbf{k}} \left( \langle \rho^*_{\mathbf{k}}(\tau) \rho_{\mathbf{k}}(\tau) \rangle_0 - 2\right) d\tau, \label{OCExprMany}\\
\langle O_0 \rangle_0 := & \frac{1}{2 \hbar^2} \sum_{\mathbf{k}} \int_0^{\hbar \beta} \int_0^{\hbar \beta} |V^{(F)}_{\mathbf{k}}|^2 \langle \rho^*_{\mathbf{k}}(\tau)\rho_{\mathbf{k}}(\sigma) \rangle_0 G_{\omegaLO}(\tau-\sigma)  d\tau  d\sigma, \label{O0ExprMany} \\
\langle \tilde{O}_2 \rangle_0 := & \frac{1}{2\hbar^2} \sum_{\mathbf{k}_1, \mathbf{k}_2} \int_0^{\hbar \beta}  \int_0^{\hbar \beta} \left|V^{(1)}_{\mathbf{k}_1,\mathbf{k}_2}\right|^2 \langle \rho^*_{\mathbf{k}_2-\mathbf{k}_1}(\tau) \rho_{\mathbf{k}_2-\mathbf{k}_1}(\sigma) \rangle_0 G_{\omegaLO}(\tau-\sigma)^2 d\tau d\sigma. \label{O2TildeExprMany}
\end{align}
Only one expectation value appears in these expressions, which is also calculated in appendix \ref{sec:AppModel}:
\begin{equation} \label{rhoExpValue}
\langle \rho_{\mathbf{k}}^*(\tau) \rho_{\mathbf{k}}(\sigma) \rangle_0 = 2 e^{-\frac{\hbar}{2m_b} k^2 \mathfrak{D}_{11}(\tau-\sigma)} + 2 \cos(\mathbf{k} \cdot \mathbf{a}) e^{-\frac{\hbar}{2m_b} k^2 \mathfrak{D}_{12}(\tau-\sigma)},
\end{equation}
where the pseudotime functions $\mathfrak{D}_{11}(\tau)$ and $\mathfrak{D}_{12}(\tau)$ can be calculated from the profile functions:
\begin{align}
\mathfrak{D}_{11}(\tau) & = \frac{4}{\hbar \beta} \sum_{n=1}^{+\infty} \frac{\sin^2\left(\frac{\omega_n \tau}{2} \right)}{\omega_n^2 A_+(\omega_n)} + \frac{4}{\hbar \beta} \sum_{n=1}^{+\infty} \frac{\sin^2\left(\frac{\omega_n \tau}{2} \right)}{\omega_n^2 A_-(\omega_n)}, \label{D11Def} \\
\mathfrak{D}_{12}(\tau) & = \frac{4}{\hbar \beta} \sum_{n=1}^{+\infty} \frac{\sin^2\left(\frac{\omega_n \tau}{2} \right)}{\omega_n^2 A_+(\omega_n)} + \frac{4}{\hbar \beta} \sum_{n=1}^{+\infty} \frac{\cos^2\left(\frac{\omega_n \tau}{2} \right)}{\omega_n^2 A_-(\omega_n)} + \frac{1}{\hbar^2 \beta^2 g_0}. \label{D12Def}
\end{align}
Using \eqref{rhoExpValue} in \eqref{OCExprMany}-\eqref{O2TildeExprMany} allows for the calculation of the interaction free energy \eqref{FintApprox}. With the following integrals \cite{Verbist1991, Houtput2021, Houtput2022}:
\begin{align}
\sum_{\mathbf{q}} \left| V^{(1)}_{\mathbf{q},\mathbf{q}-\mathbf{k}} \right|^2 & = \frac{4 \mathcal{T}_1^2}{15 \tilde{V}_0} \left| V^{(F)}_{\mathbf{k}} \right|^2, \\
\sum_{\mathbf{k}} \left| V^{(F)}_{\mathbf{k}} \right|^2 \cos(\mathbf{k} \cdot \mathbf{a}) e^{-\frac{\hbar}{2 m_b} k^2 \mathfrak{D}(\tau)} & = \frac{(\hbar \omegaLO)^2 \alpha}{\sqrt{\pi \omegaLO \mathfrak{D}(\tau)}} \chi\left( \frac{\tilde{a}}{2 \sqrt{\omegaLO \mathfrak{D}(\tau)}} \right), \\
\chi(x) & := \frac{\sqrt{\pi}}{2x} \text{Erf}(x),
\end{align}
the interaction free energy can be written in terms of the pseudotime functions $\mathfrak{D}_{11}(\tau)$ and $\mathfrak{D}_{12}(\tau)$ and the dimensionless bipolaron separation $\tilde{a} := |\mathbf{a}|/\sqrt{\frac{\hbar}{2m_b \omegaLO}}$ as follows:
\begin{small}
\begin{align}
F_{\text{int}} & \approx \frac{2 \alpha \hbar \omegaLO^{\frac{3}{2}}}{\sqrt{\pi}} \int_0^{\frac{\hbar \beta}{2}} \left[ \frac{1}{\sqrt{\mathfrak{D}_{11}(\tau)}} + \frac{1}{\sqrt{\mathfrak{D}_{12}(\tau)}} \chi\left(\frac{\tilde{a}}{2 \sqrt{\omegaLO \mathfrak{D}_{12}(\tau)}} \right) \right] \left[G_{\omegaLO}(\tau) + \frac{2 \mathcal{T}_1^2}{15\tilde{V}_0} G_{\omegaLO}(\tau)^2\right] d\tau \nonumber\\
& - \hbar \omegaLO U \sqrt{\frac{2}{\pi \omegaLO \mathfrak{D}_{12}(0)}} \chi\left(\frac{\tilde{a}}{2 \sqrt{\omegaLO \mathfrak{D}_{12}(0)}} \right). \label{FintFinal}
\end{align}
\end{small}Given an expression for the profile functions $A_{\pm}(\omega_n)$,  an upper bound for the true free energy of the bipolaron can be calculated by calculating the pseudotime functions with \eqref{D11Def}-\eqref{D12Def}, and then combining equations \eqref{FBound}, \eqref{FkinFinal}, and \eqref{FintFinal}.

\subsection{Minimization of the variational upper bound}
The variational upper bound given by \eqref{FBound} must still be minimized with respect to all possible choices for the profile functions. This can be done by calculating the derivatives $\frac{\partial F}{\partial A_+(\omega_n)}$, $\frac{\partial F}{\partial A_-(\omega_n)}$ and $\frac{\partial F}{\partial g_0}$, and setting each of them equal to zero. Since the kinetic free energy $F_{\text{kin}}$ is known exactly \eqref{FkinFinal}, the upper bound can be written as:
\begin{align}
F &\leq 2 F_{\text{free}} + \frac{1}{\beta} \text{ln}\left[V \left( \frac{m g_0}{\pi \hbar \beta^2} \right)^{\frac{3}{2}} \right] + \frac{3}{2\beta} \\
& + \frac{3}{\beta} \sum_{n=1}^{+\infty} \left\{ \text{ln}[A_+(\omega_n)] + \frac{1}{A_+(\omega_n)} - 1 + \text{ln}[A_-(\omega_n)]  + \frac{1}{A_-(\omega_n)} - 1 \right\}
- F_{\text{int}}[\mathfrak{D}_{11}(\tau),\mathfrak{D}_{12}(\tau)],  \nonumber 
\end{align}
where $F_{\text{int}}$, given by \eqref{FintFinal}, depends on the profile functions only through the functional dependence on the pseudotimes $\mathfrak{D}_{11}(\tau)$ and $\mathfrak{D}_{12}(\tau)$ on the interval $\tau \in [0, \hbar \beta/2 ]$. Therefore, the derivatives of $F_{\text{int}}$ with respect to the profile functions can be calculated using the chain rule for functional derivatives \cite{Greiner1996} and the partial derivatives $\frac{\partial \mathfrak{D}_{11}(\tau)}{\partial A_{\pm}(\omega_n)}$, $\frac{\partial \mathfrak{D}_{12}(\tau)}{\partial A_{\pm}(\omega_n)}$, and $\frac{\partial \mathfrak{D}_{12}(\tau)}{\partial g_0}$ from \eqref{D11Def}-\eqref{D12Def}.
After setting each of the derivatives equal to zero, the resulting equations can straightforwardly be solved for the profile functions $A_{\pm}(\omega_n)$. Quite elegantly, it turns out that the profile functions that minimize the energy can be written in the form of their original definitions \eqref{ApDef}-\eqref{AmDef}, but where the interaction functions $f(\tau)$ and $g(\tau)$ are given by:
\begin{align}
f(\tau) &= \frac{1}{3\hbar} \frac{\delta F_{\text{int}}}{\delta \mathfrak{D}_{11}(\tau)} = \frac{\alpha \omegaLO^3}{3 \sqrt{\pi}} \frac{1}{[\omegaLO \mathfrak{D}_{11}(\tau)]^{\frac{3}{2}}} \left( G_{\omegaLO}(\tau) + \frac{2 \mathcal{T}_1^2}{15 \tilde{V}_0} G_{\omegaLO}(\tau)^2 \right), \label{fVarSolution} \\
g(\tau) & = \frac{1}{3\hbar} \frac{\delta F_{\text{int}}}{\delta \mathfrak{D}_{12}(\tau)} = \frac{\alpha \omegaLO^3}{3 \sqrt{\pi}} \frac{e^{-\frac{\tilde{a}^2}{4 \omegaLO \mathfrak{D}_{12}(\tau)}}}{[\omegaLO \mathfrak{D}_{12}(\tau)]^{\frac{3}{2}}} \left( G_{\omegaLO}(\tau) + \frac{2 \mathcal{T}_1^2}{15 \tilde{V}_0} G_{\omegaLO}(\tau)^2 - \frac{U}{\sqrt{2}\alpha \omegaLO} \delta(\tau) \right). \label{gVarSolution} 
\end{align}
Together, the expressions for the profile functions $A_{\pm}(\omega_n)$ \eqref{ApDef}-\eqref{AmDef}, the pseudotimes $\mathfrak{D}_{11}(\tau)$ and $\mathfrak{D}_{12}(\tau)$ \eqref{D11Def}-\eqref{D12Def}, and the interaction functions \eqref{fVarSolution}-\eqref{gVarSolution} form a closed set of coupled integral equations. This set of equations can be solved numerically by iteration to obtain the profile functions, which can then be used to calculate the variational upper bound for the free energy.

In this article, we are only concerned with the ground state energy of the bipolaron, which can be found from the free energy by taking the $\beta \rightarrow +\infty$ limit. In this limit, a sum over the discrete Matsubara frequencies $\omega_n$ becomes an integral over a continuous frequency variable $\omega$ through the substitution $\frac{1}{\hbar \beta} \sum_{n} \rightarrow \frac{1}{2\pi} \int \tildeff \omega$. The profile functions are then determined by   the following integral equations, obtained from taking the $\beta \rightarrow +\infty$ limit of equations \eqref{ApDef}-\eqref{AmDef}, \eqref{D11Def}-\eqref{D12Def}, and \eqref{fVarSolution}-\eqref{gVarSolution}:
\begin{small}
\begin{align}
A_+(\omega) & = 1 + \frac{4 \alpha \omegaLO^{\frac{3}{2}}}{3 \sqrt{\pi} \omega^2} \int_0^{+\infty}  \left[\frac{\sin^2\left(\frac{\omega \tau}{2} \right)}{\mathfrak{D}_{11}(\tau)^\frac{3}{2}} + \frac{\sin^2\left(\frac{\omega \tau}{2} \right) e^{-\frac{\tilde{a}^2}{4 \omegaLO \mathfrak{D}_{12}(\tau)}}}{\mathfrak{D}_{12}(\tau)^\frac{3}{2}} \right] \left(e^{-\omegaLO \tau} + \frac{2 \mathcal{T}_1^2}{15\tilde{V}_0} e^{-2\omegaLO\tau} \right) \tildeff\tau, \label{IntegralEq1} \\
A_-(\omega) & = 1 + \frac{4 \alpha \omegaLO^{\frac{3}{2}}}{3 \sqrt{\pi} \omega^2} \int_0^{+\infty}  \left[\frac{\sin^2\left(\frac{\omega \tau}{2} \right)}{\mathfrak{D}_{11}(\tau)^\frac{3}{2}} + \frac{\cos^2\left(\frac{\omega \tau}{2} \right)  e^{-\frac{\tilde{a}^2}{4 \omegaLO \mathfrak{D}_{12}(\tau)}}}{\mathfrak{D}_{12}(\tau)^\frac{3}{2}} \right] \left(e^{-\omegaLO \tau} + \frac{2 \mathcal{T}_1^2}{15\tilde{V}_0} e^{-2\omegaLO\tau} \right) \tildeff\tau \nonumber \\
& - \frac{2 \sqrt{2} U \omegaLO^{\frac{1}{2}}}{3\sqrt{\pi} \omega^2}  \frac{e^{-\frac{\tilde{a}^2}{4 \omegaLO \mathfrak{D}_{12}(0)}}}{\mathfrak{D}_{12}(0)^{\frac{3}{2}}}, \label{IntegralEq2} \\
\mathfrak{D}_{11}(\tau) & = \frac{2}{\pi} \int_0^{+\infty} \frac{\sin^2\left(\frac{\omega \tau}{2} \right)}{\omega^2 A_+(\omega)} \tildeff\omega + \frac{2}{\pi} \int_0^{+\infty} \frac{\sin^2\left(\frac{\omega \tau}{2} \right)}{\omega^2 A_-(\omega)}  \tildeff\omega, \label{IntegralEq3} \\
\mathfrak{D}_{12}(\tau) & = \frac{2}{\pi} \int_0^{+\infty} \frac{\sin^2\left(\frac{\omega \tau}{2} \right)}{\omega^2 A_+(\omega)}  \tildeff\omega + \frac{2}{\pi} \int_0^{+\infty} \frac{\cos^2\left(\frac{\omega \tau}{2} \right)}{\omega^2 A_-(\omega)}  \tildeff\omega.  \label{IntegralEq4}
\end{align}
\end{small}Once the profile functions and pseudotimes are found by solving these integral equations, the ground state energy of the bipolaron $E_0^{(\text{bip})}$ can be calculated from the following upper bound, which is the $\beta \rightarrow +\infty$ limit of \eqref{FBound}, \eqref{FkinFinal}, and \eqref{FintFinal}:
\begin{small}
\begin{align}
 \frac{E_0^{(\text{bip})}}{\hbar \omegaLO} & \leq \frac{3}{2\pi \omegaLO} \int_0^{+\infty} \left[\text{ln}[A_+(\omega)] + \frac{1}{A_+(\omega)} - 1 + \text{ln}[A_-(\omega)] + \frac{1}{A_-(\omega)} - 1\right] \tildeff\omega \nonumber \\
& - \frac{2 \alpha \sqrt{\omegaLO}}{\sqrt{\pi}} \int_0^{+\infty} \left[
\frac{1}{\sqrt{\mathfrak{D}_{11}(\tau)}} + \frac{1}{\sqrt{\mathfrak{D}_{12}(\tau)}} \chi\left(\frac{\tilde{a}}{2 \sqrt{\omegaLO \mathfrak{D}_{12}(\tau)}} \right) \right] \left[e^{-\omegaLO \tau} + \frac{2 \mathcal{T}_1^2}{15\tilde{V}_0} e^{-2\omegaLO \tau}\right] \tildeff\tau \nonumber \\
& + U \sqrt{\frac{2}{\pi \omegaLO \mathfrak{D}_{12}(0)}} \chi\left(\frac{\tilde{a}}{2 \sqrt{\omegaLO \mathfrak{D}_{12}(0)}} \right), \label{E0VariationalFinal}
\end{align}
\end{small}In \secref{sec:Stability}, the integral equations \eqref{IntegralEq1}-\eqref{IntegralEq4} will be solved both numerically at all couplings, as well as analytically in the weak- and strong-coupling limits. In order to determine whether the bipolaron is stable, we will also need the energy of a single polaron. This result can be obtained by assuming the two electrons are infinitely far away: $\tilde{a} \rightarrow +\infty$. In that case, it holds that $A_+(\omega) = A_-(\omega) := A(\omega)$. $\mathfrak{D}_{12}(\tau)$ does not influence the results and can be dropped, and we may denote $\mathfrak{D}_{11}(\tau) := \mathfrak{D}(\tau)$. Then, equations \eqref{IntegralEq1}-\eqref{IntegralEq4} for $A(\omega)$ and $\mathfrak{D}(\tau)$ reduce to the integral equations for the single polaron problem:
\begin{align}
A(\omega) & = 1 + \frac{4 \alpha \omegaLO^{\frac{3}{2}}}{3 \sqrt{\pi} \omega^2} \int_0^{+\infty}  \frac{\sin^2\left(\frac{\omega \tau}{2} \right)}{\mathfrak{D}(\tau)^\frac{3}{2}} \left(e^{-\omegaLO \tau} + \frac{2 \mathcal{T}_1^2}{15\tilde{V}_0} e^{-2\omegaLO\tau} \right) \tildeff\tau, \label{IntegralEqOnePolaron1} \\
\mathfrak{D}(\tau) & = \frac{4}{\pi} \int_0^{+\infty} \frac{\sin^2\left(\frac{\omega \tau}{2} \right)}{\omega^2 A(\omega)} \tildeff\omega. \label{IntegralEqOnePolaron2}
\end{align}
The upper bound for the energy of a single polaron $E_0^{(1)}$ is then half of the bipolaron upper bound \eqref{E0VariationalFinal}:
\begin{equation}
\frac{E_0^{(1)}}{\hbar \omegaLO} \leq \frac{3}{2\pi \omegaLO} \int_0^{+\infty} \left[\text{ln}[A(\omega)] + \frac{1}{A(\omega)} - 1 \right] d\omega - \frac{ \alpha \sqrt{\omegaLO}}{\sqrt{\pi}} \int_0^{+\infty}
\frac{e^{-\omegaLO \tau} + \frac{2 \mathcal{T}_1^2}{15\tilde{V}_0} e^{-2\omegaLO \tau}}{\sqrt{\mathfrak{D}(\tau)}} \tildeff\tau \label{E0VariationalOnePolaron}
\end{equation}
When anharmonicity is neglected ($\mathcal{T}_1 = 0$), equations \eqref{IntegralEqOnePolaron1}-\eqref{E0VariationalOnePolaron} reduce to the results in \cite{Rosenfelder2001}. Furthermore, if the interaction functions $f(\tau)$ and $g(\tau)$ and the profile functions $A_{\pm}(\omega)$ are chosen to match the bipolaron model action in \cite{Verbist1991}, expression \eqref{E0VariationalFinal} for the bipolaron energy reduces to the known harmonic result \cite{Verbist1991}.

\section{Conditions for bipolaron stability} \label{sec:Stability}
An important question when discussing bipolarons is whether or not the bipolaron is stable. If two electrons are introduced into a material described by the Hamiltonian \eqref{HamEl}-\eqref{HamElPh}, they will only form a bipolaron if their energy $E^{(\text{bip})}_0$ is smaller than the energy $2 E_0^{(1)}$ of two single polarons that are infinitely far away ($\tilde{a} \rightarrow +\infty$). In this section, the possibility of bipolaron formation is investigated by looking for values of $U$, $\alpha$, $\mathcal{T}_1$, and $\tilde{V}_0$ where $E^{(\text{bip})}_0 < 2 E_0^{(1)}$.

The energy $E_0^{(1)}$ of a single anharmonic polaron is not known exactly. For the Fr\"ohlich polaron, the Diagrammatic Monte Carlo method \cite{Mishchenko2000, Hahn2018} provides numerically exact results for all values of $\alpha$, but this method has not yet been applied to the anharmonic Hamiltonian \eqref{HamEl}-\eqref{HamElPh}. Therefore, we will use the upper bound \eqref{E0VariationalOnePolaron} as a reference point for the single polaron energy, so that both the single polaron energy and the bipolaron energy are treated to the same level of approximation. In the harmonic case $\mathcal{T}_1 = 0$, this energy is very close to the true ground state energy \cite{Feynman1955, Mishchenko2000, Hahn2018}: it is assumed that this remains true when $\mathcal{T}_1$ is nonzero but small. This allows us to answer the question of bipolaron stability fully within the path integral formalism, just like in \cite{Verbist1991}.

\subsection{Analytical solutions of the integral equations} \label{sec:Analytical}
In general, the integral equations \eqref{IntegralEq1}-\eqref{IntegralEq4} are too complicated to be solved analytically. However, in several limiting cases it is possible to solve the equations approximately, yielding approximate bounds for the bipolaron stability region. It is assumed that $\tilde{a} = 0$ for the remainder of this section: this assumption will be motivated further in \secref{sec:Numerical}.

Firstly, we investigate the region around the unphysical boundary \eqref{PhysicalConditionAnharmonic}, i.e. $U \approx \sqrt{2} \alpha \left( 1 + \frac{\mathcal{T}_1^2}{15 \tilde{V}_0} \right)$. In this case, the longitudinal dielectric constant $\varepsilon_L(0)$ given by \eqref{EpsilonLongitudinalConstant} is very large, so that the effective interaction between the two electrons is weak. At the end of \secref{sec:Model}, it was motivated that two weakly interacting electrons corresponds to the case where the profile functions are nearly equal to each other. Since $A_-(\omega)$ should have a $1/\omega^2$ divergence, the following ansatz for the profile functions can be proposed:
\begin{align}
A_+(\omega) & \approx A(\omega) + O(\mathsf{u}), \label{ApWeak} \\
A_-(\omega) & \approx A(\omega) + A(0) \frac{\mathsf{u}^2}{\omega^2} + O(\mathsf{u}). \label{AmWeak}
\end{align}
Here $A(\omega)$ is the one-polaron function which satisfies \eqref{IntegralEqOnePolaron1}-\eqref{IntegralEqOnePolaron2}, and $\mathsf{u} \ll 1$ is a small frequency that must be determined by self-consistently solving equations \eqref{IntegralEq1}-\eqref{IntegralEq4}. It is assumed that the neglected terms $O(\mathsf{u})$ do not have a divergence as a function of $\omega$. Using these profile functions, the pseudotimes can be calculated from \eqref{IntegralEq3}-\eqref{IntegralEq4}, which yields:
\begin{align}
\mathfrak{D}_{11}(\tau) & = \frac{4}{\pi} \int_0^{+\infty} \frac{\sin^2\left(\frac{\omega \tau}{2} \right)}{\omega^2 A(\omega)} d\omega + O(\mathsf{u}) = \mathfrak{D}(\tau) + O(\mathsf{u}), \label{D11Weak} \\
\mathfrak{D}_{12}(\tau) & = \frac{2}{\pi} \int_0^{+\infty} \frac{ d\omega }{\omega^2 A(\omega) + \mathsf{u}^2 A(0)}  + O(\mathsf{u}) = \frac{1}{\mathsf{u} A(0)} + O(\mathsf{u}), \label{D12Weak}
\end{align}
where $\mathfrak{D}(\tau)$ also satisfies equations \eqref{IntegralEqOnePolaron1}-\eqref{IntegralEqOnePolaron2}. Using these pseudotimes in \eqref{IntegralEq1}-\eqref{IntegralEq2} allows for the calculation of the profile functions, which are of the form \eqref{ApWeak}-\eqref{AmWeak} but with:
\begin{equation}
\mathsf{u}^2 = \sqrt{A(0) \omegaLO} \mathsf{u}^{\frac{3}{2}} \frac{2 \sqrt{2}}{3\sqrt{\pi}} \left[\sqrt{2}\alpha \left(1 + \frac{\mathcal{T}_1^2}{15 \tilde{V}_0}\right) - U \right].
\end{equation}
If $U > \sqrt{2} \alpha \left( 1 + \frac{\mathcal{T}_1^2}{15 \tilde{V}_0} \right)$, the only possible solution to this equation is $\mathsf{u} = 0$, which yields the model action of two uncoupled polarons with total energy $2 E_0^{(1)}$. However, when $U < \sqrt{2} \alpha \left( 1 + \frac{\mathcal{T}_1^2}{15 \tilde{V}_0} \right)$, a second solution with a lower energy is possible. The energy can be found by evaluating the upper bound \eqref{E0VariationalFinal} with \eqref{ApWeak}-\eqref{D12Weak}, which yields:
\begin{equation} \label{E0BipolaronWeak}
\left\{ \begin{array}{ll}
E_0^{(\text{bip})} \leq 2 E^{(1)}_0 - \hbar \omegaLO \frac{2 A(0)}{3\pi} \left[\sqrt{2} \alpha \left(1 + \frac{\mathcal{T}_1^2}{15 \tilde{V}_0} \right) -U \right]^2 & \text{ if }  U < \sqrt{2} \alpha \left(1 + \frac{\mathcal{T}_1^2}{15 \tilde{V}_0}\right), \\
E_0^{(\text{bip})} = 2 E^{(1)}_0 & \text{ if } U \geq \sqrt{2} \alpha \left(1 + \frac{\mathcal{T}_1^2}{15 \tilde{V}_0}\right).
\end{array} \right.
\end{equation}
In the unphysical region given by $U \leq \sqrt{2} \alpha \left(1 + \frac{\mathcal{T}_1^2}{15 \tilde{V}_0}\right)$, bipolaron formation is always possible since $E_0^{(\text{bip})} < 2 E^{(1)}_0$. This behavior is also seen in the harmonic treatment of the bipolaron problem \cite{Verbist1991}. It also makes intuitive sense: it was shown in \secref{sec:PhysicalConditions} that the effective interaction between two electrons is attractive in the unphysical region, making it perfectly feasible that the electrons pair up to form a bipolaron. From here on, we will refer to this unphysical bipolaron solution as the ``attractive solution''.

In the strong coupling limit $\alpha \gg 1$, the energy functional \eqref{E0VariationalFinal} has several other local minima, which also allow for bipolaron formation when $U > \sqrt{2} \alpha \left(1 + \frac{\mathcal{T}_1^2}{15 \tilde{V}_0}\right)$. These minima can again be found by iterating the integral equations \eqref{IntegralEq1}-\eqref{IntegralEq4} once, this time starting from the following proposal for the pseudotimes:
\begin{align}
\mathfrak{D}_{11}(\tau) & \approx \frac{1}{2} \frac{1-e^{-v_1 \omegaLO \tau}}{v_1 \omegaLO} + \frac{1}{2} \frac{1-e^{-v_2 \omegaLO \tau}}{v_2 \omegaLO} + O\left( \frac{1}{v_1^2}, \frac{1}{v_2^2} \right), \label{D11ProposalLarge} \\
\mathfrak{D}_{12}(\tau) & \approx \frac{1}{2} \frac{1-e^{-v_1 \omegaLO \tau}}{v_1 \omegaLO} + \frac{1}{2} \frac{1+e^{-v_2 \omegaLO \tau}}{v_2 \omegaLO} + O\left( \frac{1}{v_1^2}, \frac{1}{v_2^2} \right). \label{D12ProposalLarge}
\end{align}
where $v_1$ and $v_2$ are large positive parameters that are yet to be determined. This proposal is inspired by the pseudotimes from the model action in \cite{Verbist1991} at large coupling. With the harmonic average of $v_1$ and $v_2$  denoted as $2/(\frac{1}{v_1}+\frac{1}{v_2}) := v$, the profile functions \eqref{IntegralEq1}-\eqref{IntegralEq2} are given by:
\begin{align}
A_+(\omega) \approx 1 & + \frac{4 \alpha}{3\sqrt\pi} v^{\frac{3}{2}} \left( \frac{1}{\frac{\omega^2}{\omegaLO^2}+1} + \frac{\mathcal{T}_1^2}{15 \tilde{V}_0} \frac{1}{\frac{\omega^2}{\omegaLO^2}+4} \right) + O\left(\frac{\alpha}{v^{\frac{3}{2}}} \right), \label{ApProposalLarge} \\
A_-(\omega) \approx 1 & + \frac{4}{3\sqrt{\pi}} \frac{\omegaLO^2}{\omega^2} \left[ \alpha \left(1+\frac{\mathcal{T}_1^2}{15 \tilde{V}_0} \right) v^{\frac{3}{2}} - \frac{U}{\sqrt{2}} v_2^{\frac{3}{2}} \right] + O\left(\alpha \sqrt{v} \right). \label{AmProposalLarge}
\end{align}
With these profile functions, the pseudotimes can be calculated with equations \eqref{IntegralEq3}-\eqref{IntegralEq4} by decomposing the integrands into partial fractions, and approximating the resulting coefficients up to highest order. The result can again be written as \eqref{D11ProposalLarge}-\eqref{D12ProposalLarge}, but where $v_1$ and $v_2$ are given by:
\begin{equation}
\left\{ \begin{array}{ll}
 \displaystyle \frac{4\alpha v^{3/2}}{3\sqrt{\pi}}\left(1+\frac{\mathcal{T}_1^2}{15\tilde{V}_0}\right) & = v_1^2, \\
 \displaystyle \frac{4\alpha v^{3/2}}{3\sqrt{\pi}}\left(1 +\frac{\mathcal{T}_1^2}{15\tilde{V}_0}\right) - \frac{2\sqrt{2} U v_2^{3/2}}{3\sqrt{\pi}} & = v_2^2.
\end{array} \right.
\end{equation}
The solutions of this system of equations represent possible local minima of the energy at strong coupling. It can be solved exactly to find two solutions:
\begin{align}
v_{1,\pm} & = \frac{16 \alpha^2}{9 \pi} \left(1+\frac{\mathcal{T}_1^2}{15\tilde{V}_0}\right)^2 (1-x_{\pm})^3, \label{v1fromWX} \\
v_{2,\pm} & = \frac{16 \alpha^2}{9 \pi} \left(1+\frac{\mathcal{T}_1^2}{15\tilde{V}_0}\right)^2 \frac{(1-x_{\pm})^4}{1+x_{\pm}}, \label{v2fromWX} \\
\text{where } x_{\pm} & := \frac{U^2 \pm U \sqrt{U^2 + 128 \alpha^2 \left(1+\frac{\mathcal{T}_1^2}{15\tilde{V}_0}\right)^2}}{64 \alpha^2 \left(1+\frac{\mathcal{T}_1^2}{15\tilde{V}_0}\right)^2}.
\end{align}
Since it is explicitly assumed that $v_1, v_2 \geq 0$, these solutions are only valid if $-1 < x_{\pm} < 1$. The ground state energy of the bipolaron can be calculated by using \eqref{D11ProposalLarge}-\eqref{AmProposalLarge} with the values \eqref{v1fromWX}-\eqref{v2fromWX} for $v_1$ and $v_2$. This yields:
\begin{equation} \label{E0BipolaronStrong}
E_{0}^{\text{(bip)}} \leq 2 E_0^{(1)} \left[ \begin{array}{l}
8 (1-x_{\pm})^2 \left(1-\frac{U}{2 (1+x_{\pm})^{\frac{1}{2}} \sqrt{2} \alpha \left(1 + \frac{\mathcal{T}_1^2}{15 \tilde{V}_0} \right)}\right) \\
\hspace{10pt} - 2(1-x_{\pm})^3 \left(1+\sqrt{1-\frac{U}{(1+x_{\pm})^{\frac{3}{2}}\sqrt{2} \alpha \left(1 + \frac{\mathcal{T}_1^2}{15 \tilde{V}_0} \right)}} \right)
 \end{array} \right],
\end{equation}
where $E_0^{(1)} = -\frac{\alpha^2}{3\pi} \left(1 + \frac{\mathcal{T}_1^2}{15 \tilde{V}_0} \right)^2$ is the energy of one polaron in the strong coupling limit \cite{Houtput2021, HoutputPhD}: this can be verified by repeating the strong coupling calculation presented here, in the case $\tilde{a} \rightarrow +\infty$. Note that $E_0^{(1)} \leq 0$: therefore, bipolaron formation is possible when the factor between square brackets in \eqref{E0BipolaronStrong} is larger than one.

\begin{figure}
\centering
\includegraphics[width=8.6cm]{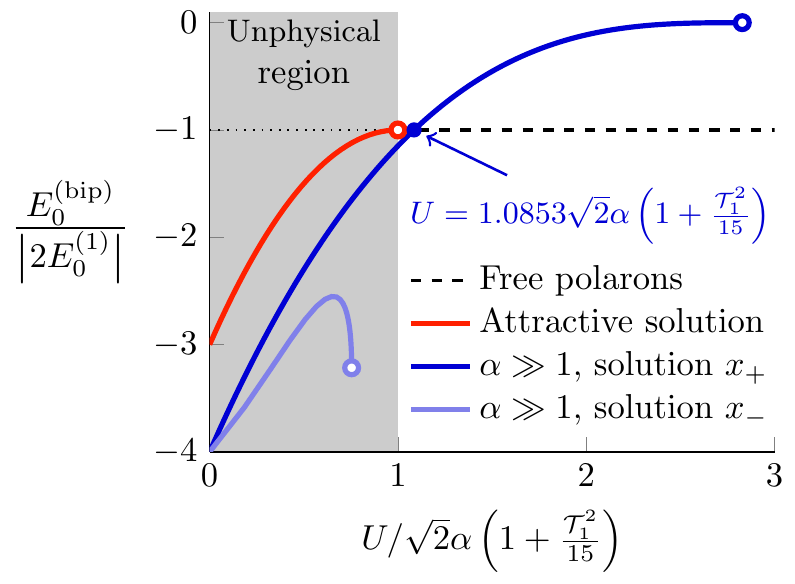}
\caption{\label{fig:StrongCoupling} The four possible local minima of the bipolaron energy functional \eqref{E0VariationalFinal}, in the strong coupling limit $\alpha \gg 1$. The attractive solution depends on the specific values of $\alpha$ and $\mathcal{T}_1$ and is plotted using $A(0) = 1$. The energies are given in equations \eqref{E0VariationalOnePolaron}, \eqref{E0BipolaronWeak}, and \eqref{E0BipolaronStrong} of the main text. Bipolaron formation is possible in the physical region, as long as $U < 1.0853 \sqrt{2} \alpha \left(1 + \frac{\mathcal{T}_1^2}{15 \tilde{V}_0} \right)$.}
\end{figure}
In the strong coupling limit, up to four local minima for the energy are possible: a minimum corresponding to two polarons infinitely far apart $\tilde{a} \rightarrow +\infty$ with total energy $2 E_0^{(1)}$, the minimum corresponding to the attractive solution with energy \eqref{E0BipolaronWeak}, and two minima that are only possible when $\alpha \gg 1$ with energies given by \eqref{E0BipolaronStrong}. \figref{fig:StrongCoupling} shows the energy of each of these minima: the bipolaron state with the lowest energy is the state that will form in practice. Outside of the unphysical region, only the strong coupling bipolaron corresponding to the $x_+$ solution of \eqref{E0BipolaronStrong} can form. Note that each of the solutions only exists in a limited range of $U$ values: the attractive solution smoothly transitions into the solution with two free polarons at the unphysical boundary $U = \sqrt{2} \alpha \left(1+\frac{\mathcal{T}_1^2}{15 \tilde{V}_0} \right)$, the solution based on $x_+$ ends where $x_+ = 1$, and the solution based on $x_-$ ends where the argument of the square root in \eqref{E0BipolaronStrong} becomes negative. The energy of this bipolaron is lower than two free polarons as long as:
\begin{equation} \label{PhaseLineStrong}
U < 1.0853 \sqrt{2} \alpha \left(1 + \frac{\mathcal{T}_1^2}{15 \tilde{V}_0} \right) - O(1).
\end{equation}
Therefore, within this model, bipolaron formation is certainly possible at sufficiently high values of $\alpha$ and $\mathcal{T}_1$. These analytical results qualitatively match the results in \cite{Verbist1991}: at weak coupling bipolaron formation is only possible in the unphysical region, but if $\alpha$ is large enough, bipolaron formation is possible in the physical region as long as $U$ satisfies \eqref{PhaseLineStrong}.

\subsection{Numerical calculation of the bipolaron phase diagram} \label{sec:Numerical}
The analytical treatment of \secref{sec:Analytical} gives results for the bipolaron stability in the weak and strong coupling limits. To find the bipolaron phase diagram for all values of $U$, $\alpha$, $\mathcal{T}_1$ and $\tilde{V}_0$, the integral equations \eqref{IntegralEq1}-\eqref{IntegralEq4} must be solved numerically, and the resulting energy \eqref{E0VariationalFinal} must be minimized with respect to the average separation $\tilde{a}$ of the electrons.

The basic idea of the numerical solution scheme is the same as the one in \cite{Rosenfelder2001}. First, the integrals over $\omega$ and $\tau$ are transformed to a finite domain by the substitutions $\omega := \tan^2(\pi\omega'/2)$ and $\tau := \tan^2(\pi \tau'/2)$. $\tilde{a}$ is also transformed to a finite domain by writing it as $\tilde{a} = 2 \tilde{a}'/(1-\tilde{a}'^2)$: then, $\omega', \tau', \tilde{a}' \in [0,1]$. Then, the functions $A_{\pm}(\omega')$, $\mathfrak{D}_{11}(\tau')$, and $\mathfrak{D}_{12}(\tau')$ are discretized on a grid of Gaussian points, and initial guesses for $A_\pm(\omega')$ and $\tilde{a}'$ are chosen. These initial guesses determine to which of the local minima in \figref{fig:StrongCoupling} the solution will converge. In practice, the energy must only be calculated close to the unphysical region and the bipolaron critical point: in this region, up to two local minima appear, which are reached by choosing $\tilde{a}' = 0$ and $\tilde{a}' = 1$ as initial guesses. For the initial guesses of the profile functions, we choose the ``weak coupling'' analytical solution \eqref{ApWeak}-\eqref{AmWeak} if $\alpha \left(1 + \frac{\mathcal{T}_1^2}{15 \tilde{V}_0} \right) < 3.5$, and the strong coupling analytical solution \eqref{ApProposalLarge}-\eqref{AmProposalLarge} otherwise. Once the initial guess is chosen, the pseudotimes $\mathfrak{D}_{11}(\tau)$ and $\mathfrak{D}_{12}(\tau)$ are calculated with \eqref{IntegralEq3}-\eqref{IntegralEq4}, and the ground state energy \eqref{E0VariationalFinal} is numerically minimized with respect to $\tilde{a}'$. This ground state energy $E_{0}^{(\text{bip})}$ is stored in memory. Then, the new profile functions $A_{\pm}(\omega)$ are calculated using \eqref{IntegralEq1}-\eqref{IntegralEq2}. This process is repeated until $E_{0}^{(\text{bip})}$ has converged up to nine significant digits. Finally, in order to ensure convergence with respect to the grid, the number of grid points is doubled, and the whole calculation is repeated using the newly calculated profile functions $A_{\pm}(\omega)$ as initial guesses, until $E_{0}^{(\text{bip})}$ has converged up to five significant digits. We note that even though the explicit minimization with respect to $\tilde{a}$ was included in this scheme, all the converged results in this article either ended on $\tilde{a}' = 0$ or $\tilde{a}' = 1$. The latter option corresponds to two free polarons with $|\mathbf{a}| \rightarrow +\infty$. This implies that if the bipolaron is stable, it always forms with $\mathbf{a} = \mathbf{0}$. This numerical result also validates \emph{a posteriori} the assumption of setting $\tilde{a} = 0$ in \secref{sec:Analytical}.

\begin{figure}
\centering
\includegraphics[width=8.6cm]{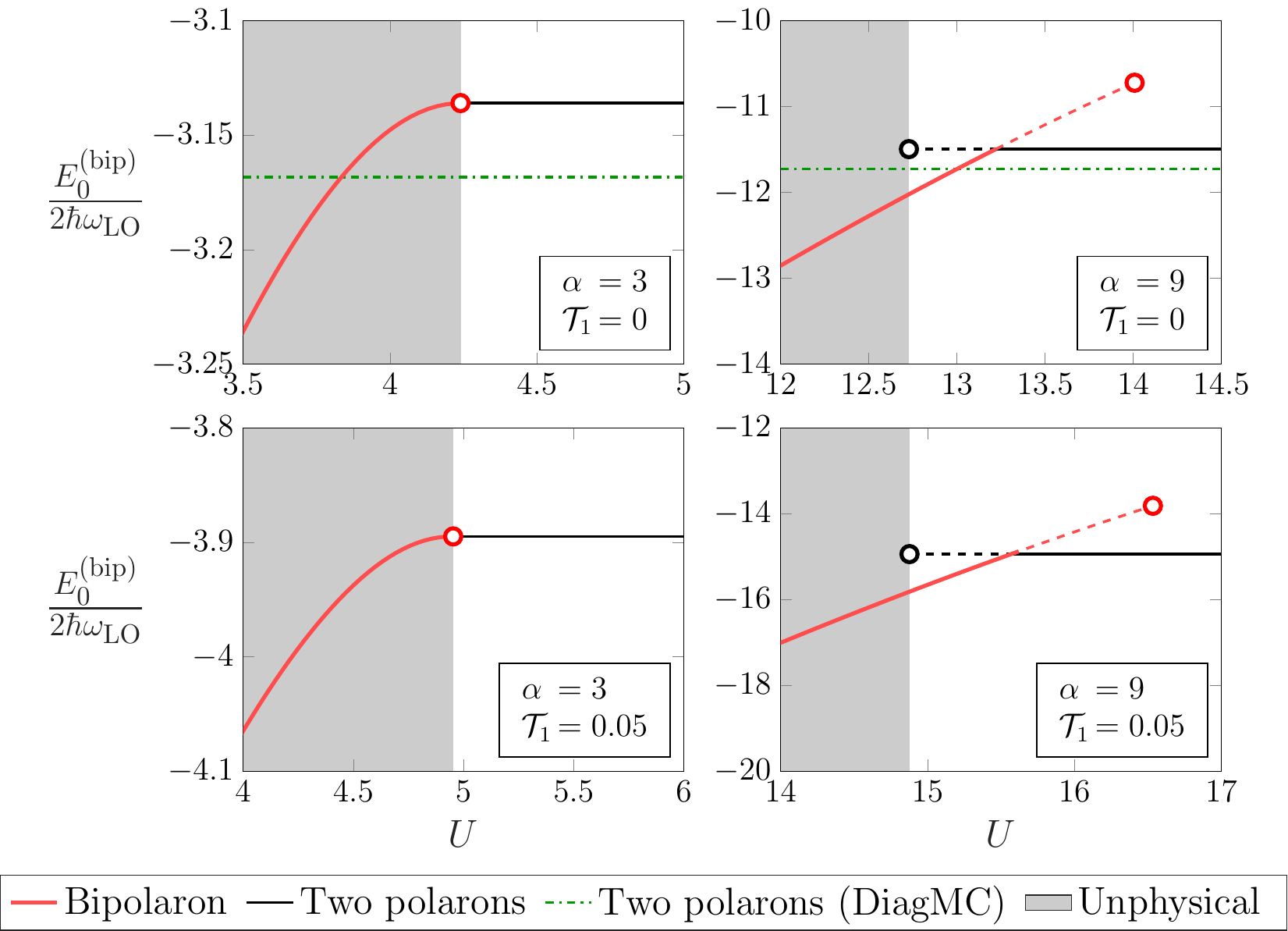}
\caption{\label{fig:BipolaronEnergies} Energy of the bipolaron as a function of $U$, for different values of $\alpha$ and $\mathcal{T}_1$. Each figure was made with $\tilde{V}_0 = 0.001$. Dashed lines represent local minima of the energy functional \eqref{E0VariationalFinal} with a higher energy than the global minimum (solid line). Dash dotted lines represent the exact energy of two free polarons, calculated with Diagrammatic Monte Carlo in the case $\mathcal{T}_1 = 0$ \cite{Mishchenko2000, Hahn2018}. When $\alpha$ and/or $\mathcal{T}_1$ are sufficiently high, bipolaron formation is possible in the physical region given by \eqref{PhysicalConditionAnharmonic}.}
\end{figure}
The above process can be repeated to calculate the bipolaron energy for different values of $U$, $\alpha$, and $\mathcal{T}_1$. \figref{fig:BipolaronEnergies} shows the bipolaron energy as a function of $U$ for several values of $\alpha$ and $\mathcal{T}_1$. Recall that this energy represents a variational upper bound, which is expected to be close to the true energy if $\mathcal{T}_1 \ll 1$ due to the approximation in \eqref{FintApprox}. In \figref{fig:BipolaronEnergies}, the harmonic results with $\mathcal{T}_1 = 0$ match both qualitatively and quantitatively with the result known in the literature \cite{Verbist1991}. When $\alpha$ is too small, bipolaron formation is impossible in the physical region. However, when $\alpha$ is large enough, there is a region where the energy functional \eqref{E0VariationalFinal} has two local minima: one corresponding to two polarons infinitely far away, and one corresponding to the bipolaron. There is a region where the bipolaron energy is lower and the physical condition \eqref{PhysicalConditionAnharmonic} is satisfied: this is where bipolaron formation is possible in practice. The energies in \cite{Verbist1991} are very close to those obtained by minimizing the general quadratic action functional \eqref{ModelAction}. Therefore, the only advantage of using the general quadratic action functional \eqref{ModelAction} is that it makes it easier to derive the analytical limits of \secref{sec:Analytical}: for the numerical minimization, the model action in \cite{Verbist1991} is sufficient. When $\mathcal{T}_1 = 0$, the bipolaron energy can also be compared to the numerically exact energy of two free polarons, calculated with Diagrammatic Monte Carlo \cite{Mishchenko2000, Hahn2018}. Since \figref{fig:BipolaronEnergies} shows a region for $\alpha = 9$ where the bipolaron energy is lower than the exact energy of two polarons, and the results in \figref{fig:BipolaronEnergies} represent an upper bound for the bipolaron energy, it is guaranteed that a physical bipolaron region exists.

\begin{figure}
\centering
\includegraphics[width=8.6cm]{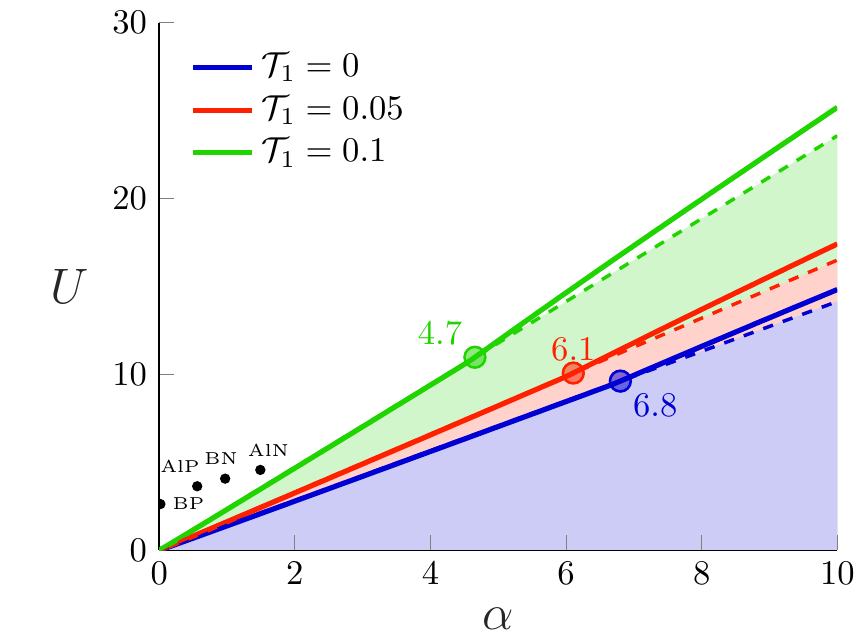}
\caption{\label{fig:PhaseDiagram} Bipolaron phase lines for different values of $\mathcal{T}_1$, where $\tilde{V}_0 = 0.001$. The full phase diagram is three-dimensional: the graph shows slices of the phase boundary at fixed values of $\mathcal{T}_1$. Bipolaron formation is possible below the solid phase line. The filled regions bounded by dashed lines are unphysical for the respective values of $\mathcal{T}_1$, in accordance to \eqref{PhysicalConditionAnharmonic}. Bipolaron formation in the physical region is only possible above a critical value $\alpha_{\text{crit}}$: the critical values are indicated on the graph with a circle. The critical value decreases as $\mathcal{T}_1$ increases. The values of $U$ and $\alpha$ of the materials in \tabref{tab:MaterialParameters} were added for clarity.}
\end{figure}
Increasing the strength of the 1-electron-2-phonon interaction $\mathcal{T}_1$ does not significantly influence the bipolaron properties when $\alpha$ is small: all it does is shift the boundary of the unphysical region to higher $U$. When $\alpha$ is large, however, the bipolaron is stable until relatively higher values of $U$. This indicates that bipolaron formation is more viable when 1-electron-2-phonon interaction is present. This becomes even more clear upon studying the bipolaron phase diagram. \figref{fig:PhaseDiagram} shows the $(U,\alpha)$ bipolaron phase diagram for different values of $\mathcal{T}_1$. The phase line is constructed as follows: for every value of $\alpha$, choose the highest value of $U$ where the bipolaron solution of \figref{fig:BipolaronEnergies} is the global minimum. Then, bipolaron formation is possible for all values $(U,\alpha)$ below the phase line: in this region the electron-phonon attraction is strong and the Coulomb repulsion is weak. The phase line exactly matches the boundary of the unphysical region \eqref{PhysicalConditionAnharmonic} until some critical value of the electron-phonon coupling $\alpha_{\text{crit}}$, above which the slope of the phase line slightly increases and a physical bipolaron region opens. Therefore, bipolaron formation is only possible when $\alpha$ is sufficiently large. This result matches the well-known harmonic result \cite{Verbist1991} and follows the qualitative prediction made from the analytical results of \secref{sec:Analytical}. Quite notably, increasing the 1-electron-2-phonon interaction significantly decreases $\alpha_{\text{crit}}$, even within the assumption that $\mathcal{T}_1$ must be small. This can be understood by noting that the Fr\"ohlich interaction and the 1-electron-2-phonon interaction both lead to an attractive interaction between the electrons, as shown in \figref{fig:ScreenedCoulomb}. When 1-electron-2-phonon interaction is present, the same net electron attraction can be reached with a weaker Fr\"ohlich interaction.

The dependence of $\alpha_{\text{crit}}$ as a function of $\mathcal{T}_1$ is shown in figure \figref{fig:alphaCritPlot}. The critical value first decreases quadratically and then linearly as a function of $\mathcal{T}_1$. Based on this behavior, the following phenomenological equation can be proposed for $\alpha_{\text{crit}}$, which fits the numerical results with an error less than $0.5 \%$ on the interval shown in \figref{fig:alphaCritPlot}:
\begin{figure}
\centering
\includegraphics[width=8.6cm]{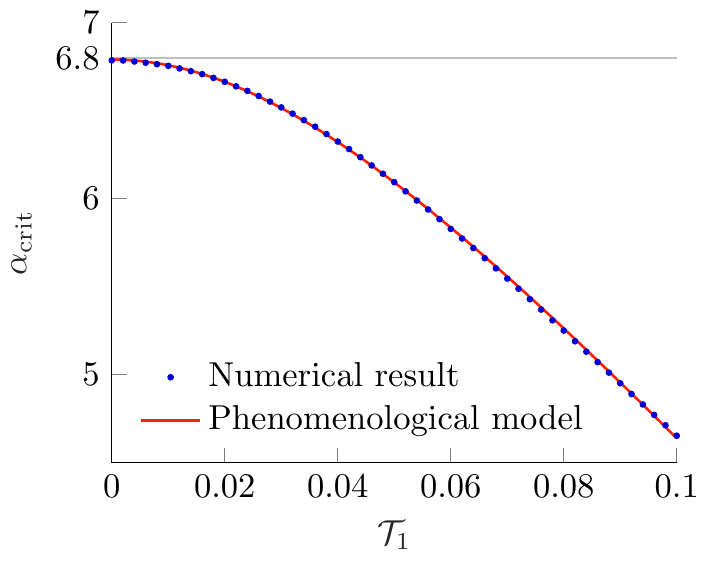}
\caption{\label{fig:alphaCritPlot} Critical value $\alpha_{\text{crit}}$ above which bipolaron formation is possible, as a function of the 1-electron-2-phonon interaction strength $\mathcal{T}_1$ where $\tilde{V}_0 = 0.001$. The critical value is significantly lowered for larger $\mathcal{T}_1$. The numerical results are fitted nicely by the phenomenological equation \eqref{PhenomenologicalCriticalValue}.}
\end{figure}
\begin{equation} \label{PhenomenologicalCriticalValue}
\alpha_{\text{crit}}(\mathcal{T}_1) \approx 6.79 - 2.10 \left(\sqrt{1 + 0.31 \frac{\mathcal{T}_1^2}{\tilde{V}_0}} - 1 \right),
\end{equation}
which contains the harmonic result $\alpha_{\text{crit}}(0) \approx 6.8$ \cite{Verbist1991} as a special case. This equation is written in terms of the combination $\mathcal{T}_1^2/\tilde{V}_0$ because in the expression for the ground state energy \eqref{E0VariationalFinal} and the integral equations \eqref{IntegralEq1}-\eqref{IntegralEq4}, the material parameters $\mathcal{T}_1$ and $\tilde{V}_0$ only appear in this combination. This is an artifact of the weak anharmonicity approximation \eqref{FintApprox}: in  general, solutions of the Hamiltonian \eqref{HamEl}-\eqref{HamElPh} can depend on $\mathcal{T}_1$ and $\tilde{V}_0$ separately \cite{HoutputPhD}. Regardless, within the theory of this article, the result must only depend on $\mathcal{T}_1^2/\tilde{V}_0$.

Within the Fr\"ohlich model \cite{Frohlich1954, Verbist1991}, three-dimensional large bipolarons are possible in theory, but the critical value $\alpha_{\text{crit}} = 6.8$ is too large in practice: materials with strong electron-phonon interaction, such as RbCl or LiF, have $\alpha \sim 4-5$ \cite{Lerner2005, LafuenteBartolome2022}. The results from this section suggest that bipolaron formation could be feasible in materials with both significant Fr\"ohlich interaction and significant 1-electron-2-phonon interaction. Indeed, even for a modest value $\mathcal{T}_1 = 0.1$, the critical value is lowered to the much more realistic value $\alpha_{\text{crit}} = 4.7$, and for materials with higher 1-electron-2-phonon interaction this critical value would be lowered even more.

\section{Discussion, outlook, and conclusion} \label{sec:Conclusions}
In this article, we investigated the energy and stability of a large bipolaron that interacts with a phonon field through both the Fr\"ohlich interaction and the 1-electron-2-phonon interaction using the path integral method \cite{Verbist1991}. In order to study the problem analytically, we used the Hamiltonian \eqref{HamEl}-\eqref{HamElPh} from \cite{Houtput2021}, where the strengths of the interactions are each determined by a single parameter: $U$ for the Coulomb interaction, $\alpha$ for the Fr\"ohlich interaction, and $\mathcal{T}_1$ for the 1-electron-2-phonon interaction. The downside of this model is that it is only applicable to a very limited range of materials.

The main result in \secref{sec:Stability} is that the 1-electron-2-phonon interaction is favorable for the stability of the large bipolaron, and that the bipolaron critical value $\alpha_{\text{crit}}$ decreases significantly as $\mathcal{T}_1$ is increased (\figref{fig:alphaCritPlot}). This suggests that large bipolaron formation may be possible in materials with significant 1-electron-2-phonon interaction. Unfortunately, the quantitative predictions in \secref{sec:Stability} are difficult to translate to real materials. As of yet, the value of $\mathcal{T}_1$ is only known for the four materials in \tabref{tab:MaterialParameters}, and these materials turned out to have negligible 1-electron-2-phonon interaction: $\mathcal{T}_1 \sim 0.001$. It is unclear whether a material exists that satisfies all the requirements \cite{Houtput2021} to use the Hamiltonian \eqref{HamEl}-\eqref{HamElPh}, and which has a significantly large value of $\mathcal{T}_1$.

Based on the results of \secref{sec:Stability}, we may expect that strontium titanate is a potential candidate for the formation of large bipolarons: it has significant Fr\"ohlich coupling to the LO phonon modes \cite{Verbist1992, Lerner2005}, but also significant 1-electron-2-phonon coupling to the soft TO phonon modes \cite{Vandermarel2019, Feigelman2021, Kumar2021}. Additionally, as a quantum paraelectric, it has a very high static dielectric constant $\varepsilon_0$ \cite{Weaver1959, Gastiasoro2020}, indicating that strontium titanate is very close to the unphysical region according to \eqref{EpsilonLongitudinalConstant}-\eqref{PhysicalConditionAnharmonic}. Whether large bipolarons can form in strontium titanate remains an open question in the literature \cite{Levstik2002, Scott2003, Levstik2003, Jourdan2003, Lin2021}. In particular, they have been proposed as an explanation for the formation of pre-formed electron pairs without a superconducting state \cite{Eagles1969, Cheng2015}, leading to a description of superconductivity in terms of a BEC-BCS crossover \cite{Hofmann2017, Gastiasoro2020}.

It is important to note that the results of this article cannot be directly applied to strontium titanate: since it has an inversion center and more than two atoms in the unit cell, the Hamiltonian \eqref{HamEl}-\eqref{HamElPh} is not valid for strontium titanate. When studying 1-electron-2-phonon interaction, the following interaction or a similar form is often used in the literature \cite{Ngai1974, Epifanov1981, Kumar2021, Kiselov2021}:
\begin{equation} \label{NgaiCoupling}
H_{2\text{ph}} = -g \int \mathbf{P}(\mathbf{r})^2 \psi^{\dagger}(\mathbf{r}) \psi(\mathbf{r}) d^3\mathbf{r},
\end{equation}
where $\psi(\mathbf{r})$ is the electron wavefunction which can be related to the electron density operator, $\mathbf{P}(\mathbf{r})$ is the polarization density which can be related to the phonon operators \cite{Kumar2021}, and $g$ is a phenomenological model parameter that determines the strength of this interaction, similarly to $\mathcal{T}_1$. This Hamiltonian is also quadratic in the phonon coordinates and can therefore be treated with the theory of \secref{sec:System} and \secref{sec:Model}. The results in \secref{sec:Stability} suggest that bipolarons are more easily stabilized when the interaction \eqref{NgaiCoupling} is included, so long as $g$ is sufficiently large. An interesting outlook is the explicit path integral treatment of bipolarons with \eqref{NgaiCoupling}, which might provide a quantitative prediction for the bipolaron phase line in strontium titanate.

We provided a semi-analytical variational upper bound for the ground state energy of the anharmonic bipolaron, and we used this upper bound to study the stability of the bipolaron in terms of the material parameters $U$, $\alpha$, and $\mathcal{T}_1$. As in the harmonic case, bipolaron formation is only possible above a critical value of $\alpha$, but this critical value strongly depends on the strength of the 1-electron-2-phonon interaction $\mathcal{T}_1$. The result can be summarized in the simple phenomenological equation \eqref{PhenomenologicalCriticalValue}, and it is shown that $\alpha_{\text{crit}} \sim 5$ for modest values of $\mathcal{T}_1$. Following these results, we suggest that large bipolaron formation is possible in materials that have strong Fr\"ohlich electron-phonon interaction as well as significant 1-electron-2-phonon interaction.

\acknowledgments

This research was funded by the University Research Fund (BOF) of the University of Antwerp (project ID: 38499). We would like to thank T. Ichmoukhamedov for many discussions on the path integral method, and S. Klimin for discussions on its application to the bipolaron problem. We also thank the research groups QMM and CMP at the University of Vienna for their collaboration on the anharmonic polaron problem, and in particular L. Ranalli, C. Verdi, C. Franchini and G. Kresse for calculating the values of $\mathcal{T}_1$ in \tabref{tab:MaterialParameters} from first principles.

\appendix

\section{Expectation values for the bipolaron model action} \label{sec:AppModel}
In this appendix, we calculate the free energy $F_0$ and several expectation values associated with the quadratic bipolaron model action \eqref{ModelAction}. The partition sum $Z_0$ of an action functional $S_0[\mathbf{r}_1(\tau),\mathbf{r}_2(\tau)]$, and the expectation value of a generic functional $A[\mathbf{r}_1(\tau),\mathbf{r}_2(\tau)]$ with respect to the action functional $S_0[\mathbf{r}_1(\tau),\mathbf{r}_2(\tau)]$, are defined as follows within the path integral formalism \cite{Kleinert2009}:
\begin{align}
Z_0 & := \int \mathcal{D}\mathbf{r}_1(\tau) \int \mathcal{D}\mathbf{r}_2(\tau) e^{-\frac{1}{\hbar} S_0[\mathbf{r}_1(\tau),\mathbf{r}_2(\tau)]}, \\
\langle A \rangle_0 & := \frac{1}{Z_0} \int \mathcal{D}\mathbf{r}_1(\tau) \int \mathcal{D}\mathbf{r}_2(\tau) A[\mathbf{r}_1(\tau),\mathbf{r}_2(\tau)] e^{-\frac{1}{\hbar} S_0[\mathbf{r}_1(\tau),\mathbf{r}_2(\tau)]},
\end{align}
where $\mathfrak{D}\mathbf{r}(\tau)$ represents a cyclic path integral. The free energy and all the expectation values that we need in this article can be found from the following path integral, which can be calculated exactly:
\begin{equation}
J[\mathbf{F}_1(\tau),\mathbf{F}_2(\tau)] = \int \mathcal{D}\mathbf{r}_1(\tau) \int \mathcal{D}\mathbf{r}_2(\tau) e^{-\frac{1}{\hbar} S_0[\mathbf{r}_1(\tau),\mathbf{r}_2(\tau)] + i \overset{2}{\underset{j=1}{\sum}} \int_0^{\hbar \beta} \mathbf{F}_j(\tau) \cdot \mathbf{r}_j(\tau)  \tildeff\tau},
\end{equation}
where $\mathbf{F}_1(\tau)$ and $\mathbf{F}_2(\tau)$ represent forces on the electrons. Once $J[\mathbf{F}_1(\tau),\mathbf{F}_2(\tau)]$ is known, the partition sum $Z_0$ and the generating expectation values of the model action can be calculated as follows:
\begin{align}
Z_0 & = J[\mathbf{0},\mathbf{0}], \\
\left\langle \exp\left(i \sum_{j=1}^2 \int_0^{\hbar \beta} \mathbf{F}_j(\tau) \cdot \mathbf{r}_j(\tau) \tildeff \tau \right) \right\rangle_0  & = \frac{J[\mathbf{F}_1(\tau),\mathbf{F}_2(\tau)]}{J[\mathbf{0},\mathbf{0}]}. \label{GeneratingExpectationValues}
\end{align}
All the expectation values in this article can be calculated from \eqref{GeneratingExpectationValues} by choosing specific expressions for $\mathbf{F}_1(\tau)$ and $\mathbf{F}_2(\tau)$.

First, note that the separation $\mathbf{a}$ can be eliminated from the path integral by the following substitutions:
\begin{align}
\mathbf{r}_1(\tau) & \rightarrow \mathbf{x}_1(\tau) + \frac{\mathbf{a}}{2}, \\
\mathbf{r}_2(\tau) & \rightarrow \mathbf{x}_2(\tau) - \frac{\mathbf{a}}{2}.
\end{align}
Then, the effect of $\mathbf{a}$ can be separated immediately:
\begin{equation} \label{TermWithA}
J[\mathbf{F}_1(\tau),\mathbf{F}_2(\tau)] = \exp\left(\frac{i\mathbf{a}}{2} \cdot \int_0^{\hbar \beta} [\mathbf{F}_1(\tau) - \mathbf{F}_2(\tau)] \tildeff \tau\right) \left. J[\mathbf{F}_1(\tau),\mathbf{F}_2(\tau)] \right|_{\mathbf{a}=\mathbf{0}}.
\end{equation}
The remaining path integral is most easily calculated in Fourier-Matsubara space. Using the conventions \eqref{rjFourierMatsubara}-\eqref{MatsubaraFrequencies} for the Fourier-Matsubara series, and using the following integration measure for the path integrals \cite{Kleinert2009}:
\begin{equation}
\int \mathcal{D} \mathbf{r}(\tau) \rightarrow \int_{\mathbb{R}^3}  \frac{\tildeff^3\mathbf{r}(\omega_0)}{\left( \frac{2 \pi \hbar^2 \beta}{m_b}\right)^{\frac{3}{2}}} \prod_{n=1}^{+\infty} \int_{\mathbb{C}^3} \frac{\tildeff^3 \mathbf{r}(\omega_n)}{\left(\frac{\pi}{\beta m_b \omega_n^2}\right)^3},
\end{equation}
a straightforward calculation yields:
\begin{align}
& \left. J[\mathbf{F}_1(\tau),\mathbf{F}_2(\tau)] \right|_{\mathbf{a}=\mathbf{0}} = \prod_{j=1}^2 \left( \int_{\mathbb{R}^3}  \frac{\tildeff^3\mathbf{r}_{j}(\omega_0)}{\left( \frac{2 \pi \hbar^2 \beta}{m_b}\right)^{\frac{3}{2}}} \prod_{n=1}^{+\infty} \int_{\mathbb{C}^3} \frac{\tildeff^3 \mathbf{r}_{j}(\omega_n)}{\left(\frac{\pi}{\beta m_b \omega_n^2}\right)^3} \right) \times \\
& \hspace{15pt} \times \exp\left( \begin{array}{l}
\underset{n \in \mathbb{Z}}{\sum} -\frac{m_b \beta \omega_n^2}{4} \left[ A_+(\omega_n) \left|\mathbf{r}_{1}(\omega_n) + \mathbf{r}_{2}(\omega_n)\right|^2 + A_-(\omega_n) \left|\mathbf{r}_{1}(\omega_n) - \mathbf{r}_{2}(\omega_n)\right|^2 \right] \\
- i \hbar \beta \underset{n \in \mathbb{Z}}{\sum} [ \mathbf{F}_{1}(\omega_n) \cdot \mathbf{r}_{1}(-\omega_n) + \mathbf{F}_{2}(\omega_n) \cdot \mathbf{r}_{2}(-\omega_n)]
\end{array}  \right), \nonumber
\end{align}
Note that the Fourier-Matsubara with $n<0$ are not independent, since $\mathbf{r}_j(-\omega_n) = \mathbf{r}^*_j(\omega_n)$: therefore, the integration measure only contains integrals over the Fourier-Matsubara components with $n \geq 0$. The above multidimensional integral can be reduced to a product of independent one-dimensional Gaussian integrals by writing the coordinates $\mathbf{r}_1$ and $\mathbf{r}_2$ in terms of the center-of-mass coordinates:
\begin{align}
\mathbf{R}(\omega_n) & = \frac{\mathbf{r}_1(\omega_n) + \mathbf{r}_2(\omega_n)}{2}, \\
\mathbf{r}(\omega_n) & = \mathbf{r}_2(\omega_n) - \mathbf{r}_1(\omega_n).
\end{align}
This transformation has a Jacobian of $1$. Then, using the fact that $A_+(0)$ is constant and $A_-(0)$ diverges as \eqref{AmLimit}, the integral becomes a product of several independent Gaussian integrals:
\begin{small}
\begin{align}
& \left. J[\mathbf{F}_1(\tau),\mathbf{F}_2(\tau)] \right|_{\mathbf{a}=\mathbf{0}}  = \int_{\mathbb{R}^3} \frac{\tildeff^3\mathbf{R}(\omega_0)}{\left( \frac{2 \pi \hbar^2 \beta}{m_b}\right)^{\frac{3}{2}}} \exp\left( - i \hbar \beta [\mathbf{F}_1(\omega_0) + \mathbf{F}_2(\omega_0)]\cdot \mathbf{R}(\omega_0) \right) \label{Integral1} \\
& \times \int_{\mathbb{R}^3} \frac{\tildeff^3\mathbf{r}(\omega_0)}{\left( \frac{2 \pi \hbar^2 \beta}{m}\right)^{\frac{3}{2}}} \exp\left( -\frac{m_b \hbar \beta^2 g_0}{2} \left|\mathbf{r}(\omega_0)\right|^2 - \frac{i \hbar \beta}{2} [\mathbf{F}_2(\omega_0) - \mathbf{F}_1(\omega_0)]\cdot \mathbf{r}(\omega_0) \right) \label{Integral2} \\
& \times \prod_{n=1}^{+\infty} \int_{\mathbb{C}^3} \frac{\tildeff^3 \mathbf{R}(\omega_n)}{\left(\frac{\pi}{\beta m_b \omega_n^2}\right)^3} \exp\left( \begin{array}{l}
 -2m_b \beta \omega_n^2 A_+(\omega_n) \left|\mathbf{R}(\omega_n)\right|^2 \\
 - i \hbar \beta \left\{ [\mathbf{F}_1(\omega_n) + \mathbf{F}_2(\omega_n)]\cdot \mathbf{R}(\omega_n) + [\mathbf{F}_1(-\omega_n) + \mathbf{F}_2(-\omega_n)]\cdot \mathbf{R}^*(\omega_n) \right\}
 \end{array}
 \right) \label{Integral3} \\
& \times \prod_{n=1}^{+\infty} \int_{\mathbb{C}^3} \frac{\tildeff^3 \mathbf{r}(\omega_n)}{\left(\frac{\pi}{\beta m_b \omega_n^2}\right)^3} \exp\left( \begin{array}{l}
 -\frac{m_b \beta \omega_n^2}{2} A_-(\omega_n) \left|\mathbf{r}(\omega_n)\right|^2 \\
 - \frac{i \hbar \beta}{2} \left\{ [\mathbf{F}_2(\omega_n) - \mathbf{F}_1(\omega_n)]\cdot \mathbf{r}(\omega_n) + [\mathbf{F}_2(-\omega_n) - \mathbf{F}_1(-\omega_n)]\cdot \mathbf{r}^*(\omega_n) \right\}
 \end{array}
 \right). \label{Integral4}
\end{align}
\end{small}
The first integral \eqref{Integral1} only converges if $\mathbf{F}_1(\omega_0) + \mathbf{F}_2(\omega_0) = \mathbf{0}$, or equivalently if:
\begin{equation} \label{FCondition}
\int_0^{\hbar \beta} [\mathbf{F}_1(\tau) + \mathbf{F}_2(\tau)] \tildeff \tau = \mathbf{0}.
\end{equation}
This condition states that the total force exerted on the model system must be zero. Whenever we will calculate expectation values in this article, this condition will be satisfied. In that case, we assume the integral over the volume $\mathbb{R}^3$ is over a large but finite volume $V$, so that $\int_{\mathbb{R}^3} \tildeff^3\mathbf{R}(\omega_0) = V$. The second integral \eqref{Integral2} converges only if $g_0 \neq 0$, which we will assume for the remainder of the derivation. The other integrals converge and can be straightforwardly calculated, since they are simple 3-dimensional Gaussian integrals. The result is:
\begin{align}
& \left. J[\mathbf{F}_1(\tau),\mathbf{F}_2(\tau)] \right|_{\mathbf{a}=\mathbf{0}}  = V \left( \frac{m_b}{2 \pi \hbar^2 \beta}\right)^{\frac{3}{2}} \left(\prod_{n = 1}^{+\infty} \frac{1}{A_+(\omega_n)} \right)^{3} \frac{1}{(2\hbar^3 \beta^3 g_0)^{\frac{3}{2}}} \left(\prod_{n = 1}^{+\infty} \frac{1}{A_-(\omega_n)} \right)^{3} \nonumber \\
& \times \exp\left( \begin{array}{l}
\displaystyle - \sum_{n = 1}^{+\infty} \frac{\hbar^2 \beta}{2m_b\omega_n^2} \left[ \begin{array}{l}
\left(\frac{1}{A_+(\omega_n)} + \frac{1}{A_-(\omega_n)}\right) \left[\mathbf{F}_{1}(\omega_n) \cdot \mathbf{F}_{1}(-\omega_n) + \mathbf{F}_{2}(\omega_n) \cdot \mathbf{F}_{2}(-\omega_n) \right] \\
 + \left(\frac{1}{A_+(\omega_n)} - \frac{1}{A_-(\omega_n)} \right) \left[\mathbf{F}_{1}(\omega_n) \cdot \mathbf{F}_{2}(-\omega_n) + \mathbf{F}_{2}(\omega_n) \cdot \mathbf{F}_{1}(-\omega_n) \right]
\end{array} \right] \\
\displaystyle - \frac{\hbar}{8 m_b g_0} \left[\mathbf{F}_1(\omega_0) - \mathbf{F}_2(\omega_0) \right]^2
\end{array} \right).
\end{align}
This expression can be written in a more intuitive form by rewriting the forces in imaginary time $\tau$ using \eqref{rjRealSpace}. This leads to the final result:
\begin{align}
Z_0 & = V \left( \frac{m_b}{2 \pi \hbar^2 \beta}\right)^{\frac{3}{2}} \left(\prod_{n = 1}^{+\infty} \frac{1}{A_+(\omega_n)} \right)^{3} \frac{1}{(2\hbar^3 \beta^3 g_0)^{\frac{3}{2}}} \left(\prod_{n = 1}^{+\infty} \frac{1}{A_-(\omega_n)} \right)^{3}, \label{PartitionSum} \\
\left. J[\mathbf{F}_1(\tau),\mathbf{F}_2(\tau)] \right|_{\mathbf{a}=\mathbf{0}} & = Z_0 \exp\left(\frac{\hbar}{4m_b} \int_0^{\hbar \beta} \int_0^{\hbar \beta} \left[ \begin{array}{l}
\mathfrak{D}_{11}(\tau-\sigma) \overset{2}{\underset{i=1}{\sum}} \mathbf{F}_{i}(\tau)\cdot\mathbf{F}_{i}(\sigma) \\
2\mathfrak{D}_{12}(\tau-\sigma) \mathbf{F}_{1}(\tau)\cdot\mathbf{F}_{2}(\sigma)
\end{array}  \right] \tildeff\tau \tildeff\sigma \right), \label{JResult}
\end{align}
where the pseudotimes are defined as:
\begin{align}
\mathfrak{D}_{11}(\tau) & = -\frac{2}{\hbar \beta} \sum_{n=1}^{+\infty} \frac{\cos(\omega_n \tau)}{\omega_n^2 A_+(\omega_n)} - \frac{2}{\hbar \beta} \sum_{n=1}^{+\infty} \frac{\cos(\omega_n \tau)}{\omega_n^2 A_-(\omega_n)} - \frac{1}{2 \hbar^2 \beta^2 g_0} + c, \label{D11Def2} \\
\mathfrak{D}_{12}(\tau) & = -\frac{2}{\hbar \beta} \sum_{n=1}^{+\infty} \frac{\cos(\omega_n \tau)}{\omega_n^2 A_+(\omega_n)} + \frac{2}{\hbar \beta} \sum_{n=1}^{+\infty} \frac{\cos(\omega_n \tau)}{\omega_n^2 A_-(\omega_n)} + \frac{1}{2 \hbar^2 \beta^2 g_0} + c. \label{D12Def2}
\end{align}
Because of condition \eqref{FCondition}, the pseudotimes are only defined up to the same constant $c$: the value of $c$ does not influence any expectation value $\langle \rangle_0$. We use this freedom to choose $c$ such that $\mathfrak{D}_{11}(0) = 0$:
\begin{equation} \label{cdef}
c = \frac{2}{\hbar \beta} \sum_{n=1}^{+\infty} \frac{1}{\omega_n^2 A_+(\omega_n)} + \frac{2}{\hbar \beta} \sum_{n=1}^{+\infty} \frac{1}{\omega_n^2 A_-(\omega_n)} + \frac{1}{2 \hbar^2 \beta^2 g_0},
\end{equation}
which simplifies further calculations. Combining \eqref{cdef} with \eqref{D11Def2}-\eqref{D11Def2} gives the expressions \eqref{D11Def}-\eqref{D12Def} for the pseudotimes that were presented in the main text.

The free energy of the model action follows immediately from the partition sum \eqref{PartitionSum}:
\begin{equation} \label{FreeEnergy2}
F_0 = \frac{3}{\beta} \sum_{n=1}^{+\infty} \left[\text{ln}(A_+(\omega_n)) + \text{ln}(A_-(\omega_n))\right] - \frac{1}{\beta} \text{ln}\left( V \left( \frac{m}{2 \pi \hbar^2 \beta}\right)^{\frac{3}{2}} \right) + \frac{3}{2\beta} \text{ln}(2 \hbar^3 \beta^3 g_0),
\end{equation}
and it is easily shown that this expression is equivalent to \eqref{FreeEnergyModel} of the main text. This expression, as well as expressions \eqref{D11Def}-\eqref{D12Def} for the pseudotimes, are in principle only valid for $g_0 \neq 0$. If $g_0 = 0$, the integral in \eqref{Integral2} must also be taken over a finite volume in order to obtain a finite result. However, we only apply the results of this appendix to the temperature zero case, $\beta \rightarrow +\infty$. A careful calculation yields that in this limit, the problematic terms with $g_0$ in \eqref{D11Def}-\eqref{D12Def} and \eqref{FreeEnergy2} all disappear, making this subtlety irrelevant.

The general expectation value \eqref{GeneratingExpectationValues} is found by combining \eqref{TermWithA} and \eqref{JResult}:
\begin{align}
& \left\langle \exp\left(i \int_0^{\hbar \beta} \left(\mathbf{F}_1(\tau) \cdot \mathbf{r}_1(\tau) + \mathbf{F}_2(\tau) \cdot \mathbf{r}_2(\tau) \right) \tildeff\tau \right) \right\rangle_0  = \exp\left(\frac{i\mathbf{a}}{2} \cdot \int_0^{\hbar \beta} [\mathbf{F}_1(\tau)-\mathbf{F}_2(\tau)] \tildeff\tau\right) \times \nonumber \\
& \times \exp\left(\frac{\hbar}{4m} \int_0^{\hbar \beta} \int_0^{\hbar \beta} \left( \begin{array}{l}
\mathfrak{D}_{11}(\tau-\sigma) [\mathbf{F}_1(\tau)\cdot\mathbf{F}_1(\sigma) + \mathbf{F}_2(\tau)\cdot\mathbf{F}_2(\sigma)] \\
 + 2 \mathfrak{D}_{12}(\tau-\sigma) \mathbf{F}_1(\tau)\cdot\mathbf{F}_2(\sigma)
\end{array} \right) \tildeff\tau \tildeff\sigma \right) \label{GeneralExpectationValue},
\end{align}
For the remaining calculations, the following four expectation values are sufficient \cite{Verbist1991}, which can be found from \eqref{GeneralExpectationValue} by taking the forces proportional to Dirac delta functions:
\begin{align}
\left\langle e^{i \mathbf{k}\cdot[\mathbf{r}_1(\tau)-\mathbf{r}_1(\sigma)]} \right\rangle_0 & = e^{- \frac{\hbar k^2}{2m_b} \mathfrak{D}_{11}(\tau-\sigma)}, & \left\langle e^{i \mathbf{k}\cdot[\mathbf{r}_2(\tau)-\mathbf{r}_2(\sigma)]} \right\rangle_0 & = e^{- \frac{\hbar k^2}{2m_b} \mathfrak{D}_{11}(\tau-\sigma)}, \label{ExpectationTwoPoint_Same} \\
\left\langle e^{i \mathbf{k}\cdot[\mathbf{r}_1(\tau)-\mathbf{r}_2(\sigma)]} \right\rangle_0 & = e^{i \mathbf{k}\cdot\mathbf{a}} e^{- \frac{\hbar k^2}{2m_b} \mathfrak{D}_{12}(\tau-\sigma)}, & \left\langle e^{i \mathbf{k}\cdot[\mathbf{r}_2(\tau)-\mathbf{r}_1(\sigma)]} \right\rangle_0 & = e^{-i \mathbf{k}\cdot\mathbf{a}}e^{- \frac{\hbar k^2}{2m_b} \mathfrak{D}_{12}(\tau-\sigma)}.\label{ExpectationTwoPoint_Different}
\end{align}
The density-density expectation value \eqref{rhoExpValue} follows immediately from these expectation values and the definition \eqref{rhoDef} of the density $\rho_{\mathbf{k}}(\tau)$.
They may also be used to calculate the expectation value \eqref{ModelActionExpValue} of the model action itself. With the following quadratic expectation values:
\begin{align}
\left\langle\left[\mathbf{r}_{j}(\tau)-\mathbf{r}_{j}(\sigma) \right]^2\right\rangle_0 & = - \left.\nabla_{\mathbf{k}}^2 \left\langle e^{i \mathbf{k}\cdot[\mathbf{r}_j(\tau)-\mathbf{r}_j(\sigma)]} \right\rangle_0 \right|_{\mathbf{k}=\mathbf{0}} = 6 \frac{\hbar}{2 m_b} \mathfrak{D}_{11}(\tau), \\
\left\langle\left[\mathbf{r}_{1}(\tau)-\mathbf{r}_{2}(\sigma) - \mathbf{a} \right]^2\right\rangle_0 & = - \left.\nabla_{\mathbf{k}}^2 \left\langle e^{i \mathbf{k}\cdot[\mathbf{r}_1(\tau)-\mathbf{r}_2(\sigma)-\mathbf{a}]} \right\rangle_0 \right|_{\mathbf{k}=\mathbf{0}} = 6 \frac{\hbar}{2 m_b} \mathfrak{D}_{12}(\tau),
\end{align}
the expectation value of the model action becomes:
\begin{small}
\begin{align}
\frac{1}{\hbar \beta} \langle S_0 - S_{\text{free}} \rangle_0 & = \frac{m_b}{2 \hbar \beta} \int_0^{\hbar \beta} \int_0^{\hbar \beta} \left(
\begin{array}{l}
\frac{1}{2} f(\tau-\sigma) \overset{2}{\underset{j=1}{\sum}} \left\langle\left[\mathbf{r}_{j}(\tau)-\mathbf{r}_{j}(\sigma) \right]^2\right\rangle_0 \\
 + g(\tau-\sigma) \left\langle\left[\mathbf{r}_{1}(\tau) - \mathbf{r}_{2}(\sigma) - \mathbf{a} \right]^2\right\rangle_0
\end{array} \right) \tildeff\tau \tildeff\sigma, \\
& = 3 \hbar \int_0^{\frac{\hbar \beta}{2}} [f(\tau) \mathfrak{D}_{11}(\tau) + g(\tau) \mathfrak{D}_{12}(\tau)] \tildeff\tau, \\
& = \frac{3}{\beta} \sum_{n=1}^{+\infty} \frac{1}{A_+(\omega_n)} \left( \frac{4}{\omega_n^2} \int_0^{\frac{\hbar \beta}{2}} \sin^2\left(\frac{\omega_n \tau}{2} \right) f(\tau) \tildeff\tau + \frac{4}{\omega_n^2} \int_0^{\frac{\hbar \beta}{2}} \sin^2\left(\frac{\omega_n \tau}{2} \right) g(\tau) \tildeff\tau \right) \nonumber \\
& + \frac{3}{\beta} \sum_{n=1}^{+\infty} \frac{1}{A_-(\omega_n)} \left( \frac{4}{\omega_n^2} \int_0^{\frac{\hbar \beta}{2}} \sin^2\left(\frac{\omega_n \tau}{2} \right) f(\tau) \tildeff\tau + \frac{4}{\omega_n^2} \int_0^{\frac{\hbar \beta}{2}} \cos^2\left(\frac{\omega_n \tau}{2} \right) g(\tau) \tildeff\tau \right) \nonumber \\
& + \frac{3}{2 \hbar \beta^2 g_0} \int_0^{\hbar \beta} g(\tau) \tildeff \tau .
\end{align}
\end{small}where in the last step we used expressions \eqref{D11Def}-\eqref{D12Def} for the pseudotimes. The integrals over $\tau$ are simply the profile functions \eqref{ApDef}-\eqref{AmDef} minus $1$. Therefore:
\begin{equation}
\frac{1}{\hbar \beta} \langle S_0 - S_{\text{free}} \rangle_0 = \frac{3}{2 \beta} + \frac{3}{\beta} \sum_{n=1}^{+\infty} \left( 1 - \frac{1}{A_+(\omega_n)} \right) + \frac{3}{\beta} \sum_{n=1}^{+\infty} \left( 1 - \frac{1}{A_-(\omega_n)} \right),
\end{equation}
which is expression \eqref{ModelActionExpValue} in the main text.

\bibliography{References}

\end{document}